\documentclass[12pt,letterpaper]{JHEP3}
\usepackage{cite}
\usepackage{epsfig}
\usepackage{amsmath}
\usepackage{amssymb}
\usepackage{amsfonts}
\usepackage{amsbsy}
\usepackage{graphicx} 
\usepackage{subfigure} 
\usepackage{mnsymbol}

\newcommand{\etan}{\eta_{\rm nuc}}
\newcommand{\Ov}[1]{\frac{1}{#1}}

\title{Boundary definition of a multiverse measure}

\author{Raphael Bousso$^{a,b,c}$, Ben Freivogel$^{a,b}$, Stefan Leichenauer$^{a,b}$, and Vladimir Rosenhaus$^{a,b}$\\ \\
  $^a$ Center for Theoretical Physics, Department of Physics\\
\ \  University of California, Berkeley, CA 94720-7300, U.S.A.\\
$^b$  Lawrence Berkeley National Laboratory, Berkeley, CA 94720-8162,
  U.S.A.\\
$^c$  Institute for the Physics and Mathematics of the Universe,\\ 
\ \ University of Tokyo,
5-1-5 Kashiwa-no-Ha, Kashiwa City, Chiba 277-8568, Japan}

\abstract{We propose to regulate the infinities of eternal inflation by relating a late time cut-off in the bulk to a short distance cut-off on the future boundary.  The light-cone time of an event is defined in terms of the volume of its future light-cone on the boundary.    We seek an intrinsic definition of boundary volumes that makes no reference to bulk structures.    This requires taming the fractal geometry of the future boundary, and lifting the ambiguity of the conformal factor.  We propose to work in the conformal frame in which the boundary Ricci scalar is constant. 

We explore this proposal in the FRW approximation for bubble universes. Remarkably, we find that the future boundary becomes a round three-sphere, with smooth metric on all scales. Our cut-off yields the same relative probabilities as a previous proposal that defined boundary volumes by projection into the bulk along timelike geodesics. Moreover, it is equivalent to an ensemble of causal patches defined without reference to bulk geodesics.  It thus yields a holographically motivated and phenomenologically successful measure for eternal inflation.}

\begin{document}

\section{Introduction}

String theory gives rise to an enormous multiverse where the constants of nature are locally constant but vary over extremely large distance scales~\cite{BP,KKLT}. In this context, it is natural to make predictions by counting. The relative probability of two events (for example, two different outcomes of a cosmological or laboratory measurement) is defined by their relative abundance,
\begin{equation}
{p_1 \over p_2} = {N_1 \over N_2}~,
\end{equation}
where $N_1$ is the expectation value of the number of times an event of type 1 occurs in the multiverse.

There is ambiguity in computing the ratio $N_1/N_2$: naively, both numbers are infinite.  Starting from finite initial conditions, eternal inflation produces an infinite spacetime volume in which everything that can happen does happen an infinite number of times.  A procedure for regulating this divergence is called a measure.  Different measures can lead to different relative probabilities starting from an otherwise identical theory.
\begin{figure}
\begin{center}
\includegraphics[scale = .7]{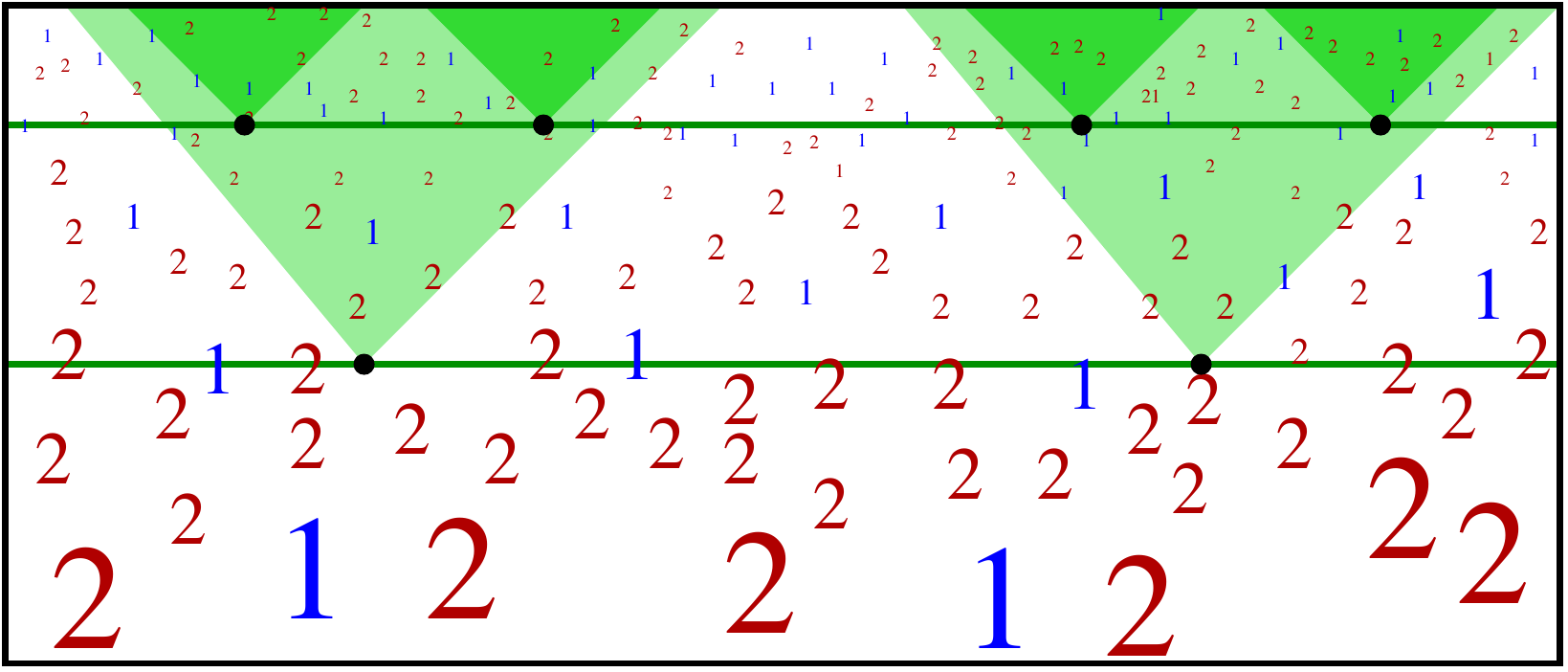}
\end{center}
\caption{Constant light-cone size on the boundary defines a hypersurface of constant ``light-cone time'' in the bulk.  The green horizontal lines show two examples of such hypersurfaces. They constitute a preferred time foliation of the multiverse. In the multiverse, there are infinitely many events of both type 1 and type 2 (say, two different values of the cosmological constant measured by observers). Their relative probability is defined by computing the ratio of the number of occurrences of each event prior to the light-cone time $t$, in the limit as $t\to \infty$.}
\label{fig-lightconecut}
\end{figure} 

A number of measures have been proposed,
including~\cite{LinMez93,LinLin94, GarLin94,GarLin94a,GarLin95,
  GarSch05,VanVil06,Van06,Bou06,Lin06,Lin07,Pag08,
  GarVil08,Win08a,Win08b,Win08c,LinVan08,Bou09}, but over the past
several years a vigorous phenomenological effort (e.g.,
\cite{GarVil05,FelHal05,PogVil06,FelHal06,SchVil06,Sch06,BouHar07,BouYan07,OluSch07,BouFre07,Sch08,BouLei08,Pag06b,HalNom07,HalSal07a,HalSal07b,DGSV08,BouFre08b,DGLNSV08,HalSal08,Sal09,BouHal09,PhiAlb09,BouLei09,HalNom09,EloGoh09})
has focused attention on a few simple proposals that remain viable.
In this paper, we will focus on the light-cone time cut-off, which
arises from an analogy with the AdS/CFT correspondence.  Given a
short-distance cut-off on the boundary conformal field theory, the
radial position of the corresponding bulk cut-off~\cite{SusWit98} can
be obtained from causality alone~\cite{BouRan01}, without reference to
the details of the bulk-boundary correspondence.  In the multiverse,
the time of a bulk event can similarly be defined in terms of a scale
on the future boundary of the multiverse~\cite{GarVil08}.  The
simplest causal relation is to associate to each bulk event a boundary
scale given by the volume of its future light-cone on the
boundary~\cite{Bou09} (see Fig.~\ref{fig-lightconecut}).\footnote{A
  different bulk-boundary relation was proposed in
  Ref.~\cite{GarVil08}.  One of us (RB) has argued that this relation
  is less well-defined than light-cone time and is not analogous to
  the UV/IR relation of AdS/CFT~\cite{Bou09}.  These concerns aside,
  it could be combined with the metric we construct on the future
  boundary.  The resulting measure would be different from the one
  obtained here.}

There is a remaining ambiguity: how is the volume of the future
light-cone to be defined?  Near the future boundary, physical
distances diverge in inflating regions and go to zero near
singularities.  
Ref.~\cite{Bou09} defined boundary volumes by erecting a congruence of geodesics
orthogonal to a fixed, fiducial bulk hypersurface and projecting
future infinity onto the fiducial hypersurface along the geodesics. The definition of Ref.~\cite{Bou09} is quite
robust\footnote{In this respect, the definition of volumes explored
  here remains inferior, for now, since it is completely well-defined
  only for homogeneous bubble universes.  However, for the reasons
  stated below, it may ultimately prove to be more fundamental.}; in
particular, it does not matter whether geodesics expand or collapse or
cross.

We would like to find an alternative definition of the boundary volume
which does not rely on such an elaborate bulk construction. The
geodesics used to project the light-cone onto the fiducial
hypersurface do only one thing for us: they define boundary volumes.
Yet they encode an enormous amount of geometric bulk information (the
exact path of each geodesic, the expansion and shear of nearby
geodesics, etc.), most of which is never used for any purpose.  They
bear no apparent relation to any physical system; for example, they do
not represent the worldlines of actual particles.  Moreover, the
construction takes an absurdly classical viewpoint of the bulk
geometry: because of the exponential expansion of de~Sitter vacua, the
projection of a late time light-cone onto the fiducial hypersurface
has subplanckian volume.

There is another reason why it would be nice to eliminate bulk
geodesics from the definition of the measure.  Ultimately, one expects
that a fundamental description of the multiverse will involve its
boundary structure in some way. This was a key motivation for seeking
a multiverse analogue of the UV/IR relation of AdS/CFT in the first
place.  The equivalence of the causal patch and light-cone time
cut-off, along with their phenomenological successes, encourages us to
take seriously the motivations behind the two measures.  In
particular, we would like to define the light-cone time cut-off, to
the greatest extent possible, in terms of quantities that are
intrinsic to the boundary.

In this paper we will propose an intrinsic definition of volumes on the future boundary of the multiverse, and we will explore the resulting time foliation and measure.  Like in AdS, we will render the boundary finite by a conformal rescaling of the bulk metric.  However, we face two difficulties that have no direct analogue in AdS/CFT.
\paragraph{The boundary of the multiverse is naturally a fractal.}  In the bulk, bubbles of different vacua keep being produced at later and later times, leading to boundary features on arbitrarily small scales.  If the bulk evolution exhibits attractor behavior (this is common to many global foliations of the multiverse, though the attractor itself depends on the foliation), then the boundary will exhibit self-similarity in the short-distance limit.  Because of the abundance of short-distance structure, the boundary must be constructed first on large scales, and then refined to better and better resolution, corresponding to evolving the bulk to later and later times.  The details of this procedure have not been carefully formulated and will concern us greatly.  

The fractal structure leads to a number of difficulties for defining a metric on the future boundary:
\begin{figure}\label{fig-onebubblemultiverse}
\begin{center}
\includegraphics[scale = .35]{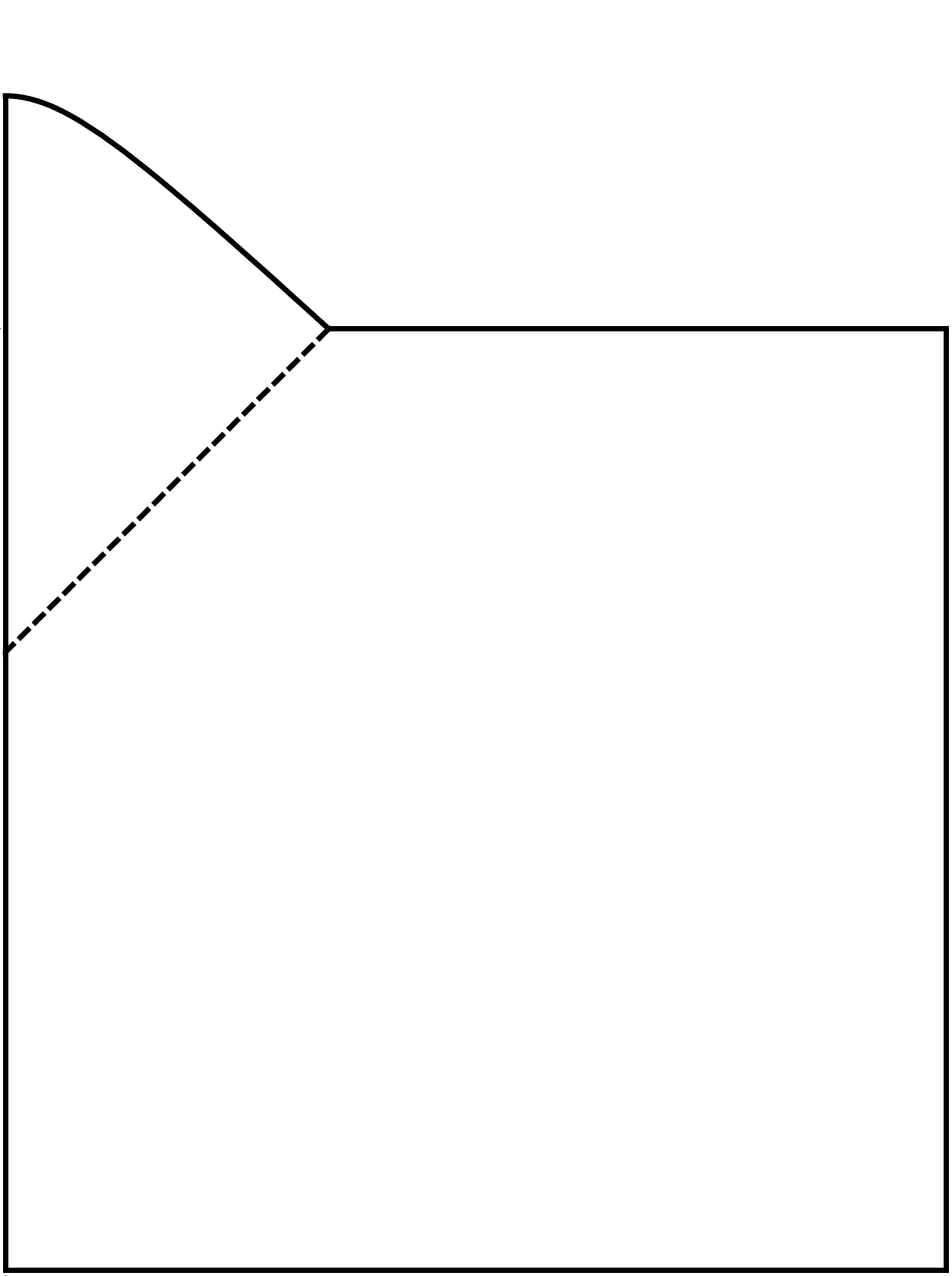}
\end{center}
\caption{Conformal diagram of de~Sitter space with a single bubble nucleation.  The parent de~Sitter space is separated from the daughter universe by a domain wall, here approximated as a light-cone with zero initial radius (dashed line).  There is a kink in the diagram where the domain wall meets future infinity.  This diagram represents a portion of the Einstein static universe, but the Ricci scalar of the boundary metric is not constant.}
\end{figure} 
\begin{itemize}

\item Generically, when a new bubble forms at late times in the bulk, one would like to include this information on the boundary, for example by coloring a small disk on the boundary which is enclosed by this bubble.  But the new vacuum will also change the metric in the bulk, and thus, the shape of the conformal diagram.  In other words, generically, the boundary {\em metric\/} will keep changing as we increase the resolution.  This is worse than the behavior of most fractals we are familiar with, which are defined by coloring points on a {\em fixed} background metric.

\item A natural way of constructing a conformal diagram depicting the formation of a bubble universe in de~Sitter space leads to a future boundary with a ``sharp edge'' (see Fig.~2), where the first derivative of the induced boundary metric is discontinuous.  With more and more bubbles forming at late times, such edges would appear on arbitrarily small scales.

\end{itemize}

\paragraph{The shape of the boundary is defined only up to finite conformal transformations.}  Because the boundary metric is obtained from the bulk spacetime by a conformal rescaling, it is only defined up to a finite conformal factor. The shape of many conformal diagrams can be changed by a conformal transformation.  The resulting diagram is just as legitimate as the original one, but its boundaries may have a different geometry.  

\begin{itemize}
\item In the multiverse, this ambiguity in the boundary metric leads to an ambiguity in the bulk foliation.  Unlike in AdS, which is asymptotically empty, this leads to a potential ambiguity in the measure.  For example, a different choice of conformal factor may change the predicted probability distribution for the cosmological constant.
\end{itemize}

\paragraph{Outline} 
We introduce the concept of light-cone time in Sec.~\ref{sec-lct}, emphasizing that it requires a definition of the volume of any future light-cone.  In Sec.~\ref{sec-ricci}, we propose that the volume should be defined as the volume enclosed by the light-cone on the future conformal boundary of the multiverse.  Moreover, we propose choosing the conformal factor so that the boundary metric has constant scalar curvature,
\begin{equation}
R = {\rm constant}~,
\label{eq-rcon}
\end{equation}
where $R$ is the Ricci scalar of the boundary metric\footnote{Garriga
  and Vilenkin \cite{GarVil09} defined the metric on future infinity
  by foliating the spacetime  by surfaces whose induced metric has $R
  = 0$. In general
  spacetimes this is different from our proposal, and bulk $R=0$
  surfaces do not always exist. However, in the special case we focus
  on of homogeneous de Sitter bubbles, their boundary metric is the
  same as ours.}  A recent result in mathematics---the solution to the Yamabe problem---guarantees that a suitable conformal transformation can always be found and is essentially unique on the boundary.  This proposal defines what we shall refer to as ``new light-cone time''.  In Sec.~\ref{sec-frw}, we apply our proposal to a simplified landscape model.  We approximate bubble universes as homogeneous open FRW cosmologies, we assume that they all have nonzero cosmological constant, and we neglect bubble collisions.  Despite these simplifications, all of the difficulties listed above arise in this model, and we show that our proposal succeeds in addressing them.  We construct the conformal diagram iteratively, making sure that the boundary condition $R=$ const.\ is satisfied each time a new bubble universe is included in the bulk. We find that this procedure leads to a fixed, everywhere smooth boundary metric that can be taken to be a unit three-sphere.   

The boundary metric picked out by the condition $R=$ const.\ defines new light-cone time.  
In Sec.~\ref{sec-prop}, we present some of the most important properties of this time foliation.  In Sec.~\ref{sec-rate}, we derive rate equations that govern the distribution of vacua.    We emphasize the boundary viewpoint, in which bubble universes correspond to topological disks on the boundary three-sphere. In Sec.~\ref{sec-asol}, we find the solution of the rate equations.  The late time attractor behavior in the bulk corresponds to a universal ultraviolet scaling of the distribution of disks on the boundary.  In Sec.~\ref{sec-count}, we derive the crucial expression that underlies the probability measure: the number of events of arbitrary type, as a function of light-cone time.

In Sec.~\ref{sec-pm}, we analyze the probability measure.  In Sec.~\ref{sec-gloglo}, we show that new light-cone time yields the same measure as old light-cone time, i.e., both cut-offs predict the same relative probabilities for any two types of events in our simplified landscape model.  In particular, this implies that our ``new'' light-cone time shares the phenomenological successes of the old one, and of the causal patch measure dual to it.

In Sec.~\ref{sec-general}, we discuss how our approach may extend to the general case, where inhomogeneities, collisions, and vacua with $\Lambda=0$ are included.  We also consider the (likely) possibility that the landscape contains vacua of different dimensionality.  In defining a unique boundary metric, several difficulties arise in addition to the ones listed above, and a more general method of implementing Eq.~(\ref{eq-rcon}) is needed.  We discuss what phenomenological properties may be expected of the resulting measure.  In particular, we expect that in the context of inhomogeneous universes, a boundary definition of light-cone volume will address a problem pointed out by Phillips and Albrecht~\cite{PhiAlb09}.

\section{New light-cone time}
\label{sec-nlt}

In this section, we define new light-cone time, and we construct the surfaces of constant light-cone time in a simple multiverse.

\subsection{Probabilities from light-cone time}
\label{sec-lct}

Given a time foliation of the multiverse, the relative probability of events of type $A$ and $B$ (e.g., two different outcomes of an experiment) can be defined by
\begin{equation}
\frac{p_A}{p_B}=\lim_{t\to \infty}\frac{N_A(t)}{N_B(t)}~,
\label{eq-p}
\end{equation}
where $N_A(t)$ is be the number of times an event of type $A$ has occurred prior to the time $t$.  This measure depends strongly on the choice of $t$.  In this paper, we focus exclusively on the case where $t$ is of the form
\begin{equation}
t\equiv -\frac{1}{3}\log \frac{V(E)}{4\pi/3}~.
\label{eq-t}
\end{equation}
Here, $V(E)$ is the (suitably defined) volume of the causal future of $E$, $I^+(E)$. Any time variable of this form will be referred to as light-cone time.

To complete the definition of light-cone time, and thus of the measure, one must define $V(E)$.  One such definition, which results in what we shall refer to as {\em old light-cone time}, was given in Ref.~\cite{Bou09}: consider a family of geodesics orthogonal to a fiducial hypersurface in the bulk, and let $V(E)$ be the volume occupied on the fiducial hypersurface by the geodesics that eventually enter $I^+(E)$.  In this paper we will explore a different definition, {\em new light-cone time}.

Often, we will find it convenient to work instead with the variable
\begin{equation}
\eta=e^{-t}=\left(\frac{V(E)}{4\pi/3}\right)^{1/3}~,
\label{eq-etadef}
\end{equation}
which is naturally interpreted as a boundary distance scale. Since hypersurfaces of constant $\eta$ are hypersurfaces of constant $t$, they define the same bulk foliation, and thus, the same measure.  In the bulk, $\eta$ decreases towards the future and vanishes on the boundary; $t$ increases towards the future and diverges at the boundary.

\subsection{A gauge choice for the conformal boundary}
\label{sec-ricci}

Physical distances diverge in the asymptotic future of an eternally inflating multiverse, except inside black holes and vacua with negative cosmological constant, where they approach zero.  Such behavior is found in many other spacetimes, such as (Anti-)de~Sitter space or the Schwarzschild solution.  However, in many cases a boundary of finite volume can be defined by a conformal transformation.

To a physical spacetime $M$, with metric $g_{\mu\nu}$, we associate an unphysical spacetime $\tilde{M}$ with metric $\tilde{g}_{\mu\nu}$ which satisfies a number of conditions.
There must exist a conformal isometry $\psi:M\rightarrow \psi[M] \subset \tilde{M}$ such that
\begin{equation}\tilde{g}_{\mu \nu} = \Omega^2(x) \psi^* g_{\mu \nu} =  e^{2\phi(x)}\psi^* g_{\mu \nu}
\end{equation} 
in $\psi[M]$.
Note that $\Omega$ should be nowhere-vanishing on $\psi[M]$, and we demand that $\Omega$ be sufficiently smooth (say, $C^3$~\cite{HawEll}) in $\psi[M]$ and extend to a continuous function on the closure, $\overline{\psi[M]}$.
We will refer to both $\Omega$ and $\phi= \log \Omega$ as the conformal factor.
Hereafter we will identify $M$ with its image $\psi[M]$ (called ``the bulk"), and refer to the physical ($g_{\mu\nu}$) and unphysical ($\tilde{g}_{\mu\nu}= \Omega^2g_{\mu\nu}$) metrics defined on $M$, eliminating $\psi$ from the notation.

As a set, the boundary of $M$ is defined as those points in the closure of $M \subset \tilde{M}$ which are not contained in $M$ itself: $\partial M  = \overline{M}-M$.
By the ``boundary metric", $G_{a b}$, we mean the unphysical metric induced on $\partial M$ viewed as a subset of $\tilde M$.
The key property we will require of the unphysical spacetime is that for any $E\subset M$, the volume of $I^+(E) \cap \partial M$ be finite in the boundary metric.  (Below we will define $V(E)$ to be this volume.)
Note that for those cases where the physical volumes and distances diverge as one approaches the boundary, this implies that the conformal factor $\Omega$ approaches zero on $\partial M$, or that $\phi$ diverges.
We will return in Sec.~\ref{sec-general} to the question of which bulk metrics can be conformally rescaled so that the boundary metric is finite and nonsingular.

There is an ambiguity in the boundary metric which is related to an ambiguity in choosing the conformal factor that makes the rescaled bulk spacetime finite. Consider an additional Weyl rescaling with a conformal factor that is {\em bounded\/} on the boundary:
\begin{equation}
\tilde G_{ab} = e^{2 \tilde \phi(x)} G_{ab} ~,
\end{equation}
with $|\phi(x^\mu)|<K$ for some real number $K$.  This will produce a new unphysical spacetime whose boundary still has finite volume, and which we could have obtained directly from the original spacetime via the conformal factor $\phi+\tilde \phi$.  So the requirement of finiteness is not sufficient to fix the boundary metric completely.

We propose to fix the ambiguity by demanding that the conformal transformation yield a boundary metric of constant scalar curvature:
\begin{equation}
R = {\rm constant}~.
\label{eq-ric}
\end{equation}
Here, $R$ is the Ricci scalar computed in the boundary metric, not the bulk Ricci scalar restricted to the boundary.  The value of the constant is arbitrary.  (It can be changed by an overall rescaling of the unphysical spacetime; this shifts light-cone time by a constant but leaves the foliation, and thus the probability measure, invariant.)  At a naive level of counting degrees of freedom, these conditions fix the metric.

At a more refined level, the question of whether an arbitrary metric
on a smooth, compact manifold can be brought to a metric with constant
scalar curvature is a difficult problem in mathematics known as the
Yamabe problem. The Yamabe problem has been solved in the affirmative ~\cite{Scho84}, ~\cite{Lee87}. For a smooth compact Riemannian manifold $(M_n, G')$ of dimension $n\geq 3$, there exists a metric $G$ conformal to $G'$ such that the Ricci scalar of $G$ is constant. The problem of finding the appropriate conformal transformation $G=\phi^{4/(n-2)} G'$ amounts to solving the differential equation
\begin{equation} \label{eq:Yamabe}
\frac{4(n-1)}{n-2} \nabla^2 \phi + R' \phi = R \phi^{\frac{n+2}{n-2}}~,
\end{equation}
with $R$ a constant.  This is nontrivial since the solution $\phi$ must be smooth and strictly positive, and must exist globally.

Yamabe ~\cite{Yam60} attempted to solve the problem by noting that (\ref{eq:Yamabe}) can be written as the Euler-Lagrange equation for a certain functional of the metric. This functional is closely related to the average scalar curvature of the metric over the manifold. Since the minimum occurs for the metric with constant curvature, the solution focuses on showing that the minimum of this functional is realized. Yamabe's original proof was later shown to be valid only in certain cases ~\cite{Tru68}. The proof was extended to other cases in ~\cite{Aub76} and completed in 1984 by Richard Schoen ~\cite{Scho84}. A unified and self-contained proof of the Yamabe problem is given in ~\cite{Lee87}.

Having shown that a constant curvature metric exists, can we be sure that it is unique?  Generically it is~\cite{And05}: there is an open and dense set $U$ in the space of conformal classes of metrics such that each element $[G] \in U$ has a unique unit volume metric with constant scalar curvature.  In simple cases with symmetries, however, there may be a few-parameter family of solutions to the Yamabe problem. For example, the round three-sphere has an ambiguity given by the global conformal group $SO(4,1)$.  However, this ambiguity can only be used to fix the locations of a few points. We will ultimately be interested in the behavior on the shortest scales where this ambiguity has no effect.

We can now define $V(E)$ in Eq.~(\ref{eq-t}) as follows.  Let $L(E)$ be the portion of the boundary in the causal future of $E$, i.e., the intersection of $I^+(E)$ with the boundary.  Let $V(E)$ be the volume of $L(E)$, measured using the metric $G_{ab}$ obtained by the (essentially unique) conformal transformation that achieves constant Ricci scalar on the boundary:
\begin{equation}
V(E)\equiv \int_{L(E)} d^3 y \sqrt{G} ~,
\end{equation}
where $G$ is the determinant of $G_{ab}$.  With this choice of $V$, Eq.~(\ref{eq-t}) defines the {\em new light-cone time}, $t$, and Eq.~(\ref{eq-p}) defines a probability measure for the multiverse, the new light-cone time cut-off.

\subsection{The multiverse in Ricci gauge}
\label{sec-frw}

In this section, we carry out the construction of future infinity with $R=$ const, subject to the following approximations and assumptions:
\begin{itemize}
\item All metastable de~Sitter vacua are long-lived: $\kappa_\alpha\ll 1$, where $\kappa_\alpha=\sum \kappa_{i\alpha}$ is the total dimensionless decay rate of vacuum $\alpha$, and $\kappa_{i\alpha}=\Gamma_{i\alpha} H_\alpha^{-4}$, where $\Gamma_{i\alpha}$ is the decay rate from the de~Sitter vacuum $\alpha$ to the arbitrary vacuum $i$ per unit four volume, and $H_\alpha=(3/\Lambda_\alpha)^{1/2}$ is the Hubble constant associated with the asymptotic de~Sitter regime of vacuum $\alpha$.\footnote{We denote vacua with $\Lambda>0$ by indices $\alpha, \beta, \ldots$ and vacua with $\Lambda\leq 0$ by indices $m,n,\ldots$.  If no particular sign of $\Lambda$ is implied, vacua are denoted by $i,j,\ldots$.}

\item Bubble collisions are neglected.  This is equivalent to setting all decay rates to zero in the causal past of a bubble.

\item Each bubble is a homogeneous, isotropic, open universe; the only exception are bubbles nucleating inside bubbles, which spontaneously break this symmetry.  On a large scale, this assumption essentially follows from the previous two approximations, since a large suppression of the decay implies a large suppression of fluctuations that break the $SO(3,1)$ symmetry of the Coleman-de Luccia geometry.  But on a small scale, this assumption implies that we suppress any structure formation that results from small initial perturbations; we treat the bubble universe as completely homogeneous and isotropic.

\item All bubbles have the same spacetime dimension, $D=3+1$.

\item No vacua with $\Lambda=0$ are produced.  This ensures that the boundary contains no null portions or ``hats'', where the boundary metric would be degenerate.

\end{itemize} 
The above assumptions allow us to take an iterative approach to the construction of the conformal diagram, while satisfying the gauge condition (\ref{eq-rcon}).  We will now describe this construction step by step.

The universe begins in some metastable vacuum (it will not matter which one).   Our initial step is to construct a conformal diagram satisfying $R=$ const.\ for this de~Sitter vacuum, {\em as if it were completely stable}.   We will refer to this spacetime as the zero-bubble multiverse.  The metric is
\begin{equation}
ds_0^2 =\frac{- d \eta^2 + d\Omega_3^2}{H_0^2 \sin^2 \eta}~,
\label{eq-dsmetric}
\end{equation}
where $d\Omega_3^2=d\xi^2+\sin^2\xi (d\theta^2+\sin^2\theta d\phi^2)$ is the metric on the unit three-sphere.  (Since an infinitely old metastable de~Sitter space has zero probability, let us restrict to $\eta\leq \pi/2$,\footnote{In naming the time coordinate in Eq.~(\ref{eq-dsmetric}) $\eta$, we are anticipating the result below that $\eta=$ constant will define hypersurfaces of constant light-cone time in accordinance with the definition of $\eta$ in Eq.~(\ref{eq-etadef}).  Since this definition requires $\eta$ to take positive values which {\em decrease} towards the future boundary, we shall take $\eta$ to have a positive range in Eq.~(\ref{eq-dsmetric}).  Thus, $\eta\geq \pi/2$ corresponds to times {\em after} $\eta=\pi/2$.} as if the universe came into being at the time $\eta=\pi/2$.)  Multiplying this metric by the conformal factor $e^{2\phi}=H_0^2\sin^2\eta$, one obtains a conformally rescaled metric 
\begin{equation}
d\tilde s_0^2 =- d \eta^2 + d\Omega_3^2~,
\label{eq-dsmetric1}
\end{equation}
which can be viewed as a portion of an Einstein static universe~\cite{HawEll}.  The future boundary corresponds to the points with $\eta=0$, which were not part of the physical spacetime.  Its induced metric is that of a round three-sphere with unit radius.  Thus, it satisfies the gauge condition we impose, that the three-dimensional Ricci scalar be constant.

The second step is to construct light-cone time in the $1$-bubble multiverse.  Consider the causal future $I^+$ of an arbitrary event $E$ with coordinates $(\eta,\xi,\theta,\phi)$ in the physical spacetime.  We note that conformal transformations map light-cones to light-cones, and that the unphysical metric, Eq.~(\ref{eq-dsmetric1}), is spatially homogeneous.  Therefore, the volume $V$ occupied by $I^+(E)$ on the future boundary can only depend on the $\eta$ coordinate of the event $E$, and not on its spatial coordinates; it is given by 
\begin{equation}
V=\int_0^\eta 4\pi \sin^2\eta d\eta~.
\end{equation}
The light-cone time is $t\equiv -\frac{1}{3}\log \frac{V}{4\pi/3}$.  At late times, $V=\frac{4\pi}{3}\eta^3$, so $t= -\log\eta$.

The next step is to consider the first bubble that nucleates inside this vacuum, i.e., the nucleation event with the earliest light-cone time.  (The time and place of this nucleation is of course governed by quantum chance; as we add more and more bubbles, we shall ensure that on average, their nucleation rate per unit four-volume conforms to the rates $\Gamma_{\alpha m}$ dictated by the landscape potential.)  We then replace the causal future of the nucleation event with the new bubble, i.e., with an open FRW universe, which may have positive or negative cosmological constant, and may contain matter and radiation.  Aside from this one nucleation event, we treat both the parent and the daughter vacuum as completely stable, so we do not modify the spacetime in any other way.  We thus obtain a physical spacetime consisting of a parent de~Sitter vacuum and one bubble universe with a different vacuum, which we refer to as the one bubble multiverse.  Now we construct a new conformal diagram for this spacetime, as follows (see Fig.~\ref{fig-cutandpaste}):

\begin{figure}[h]\label{fig-cutandpaste}\begin{center}
\subfigure[]{\includegraphics[scale = .3]{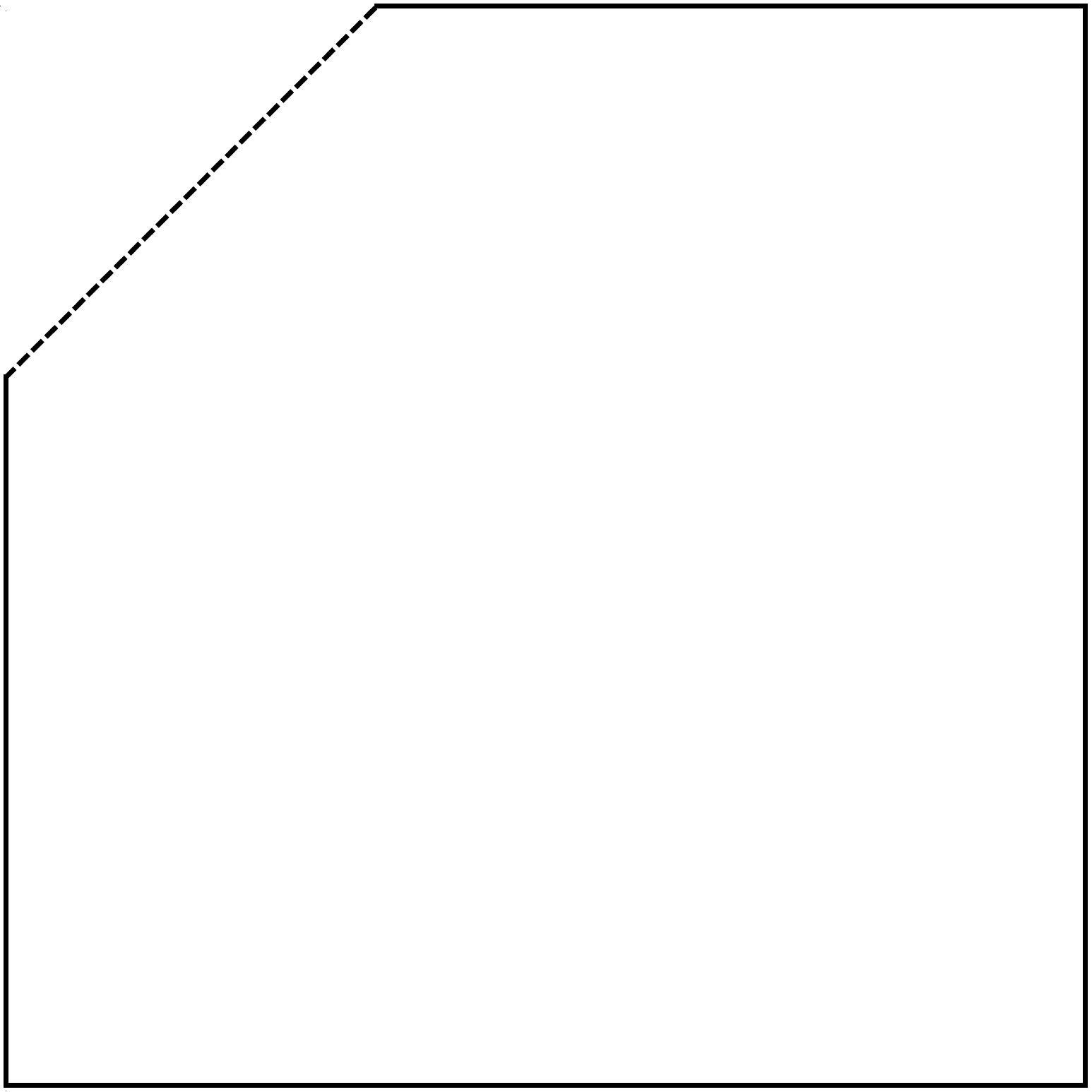}}\hspace{0.5 in}
\subfigure[]{\includegraphics[scale = .5]{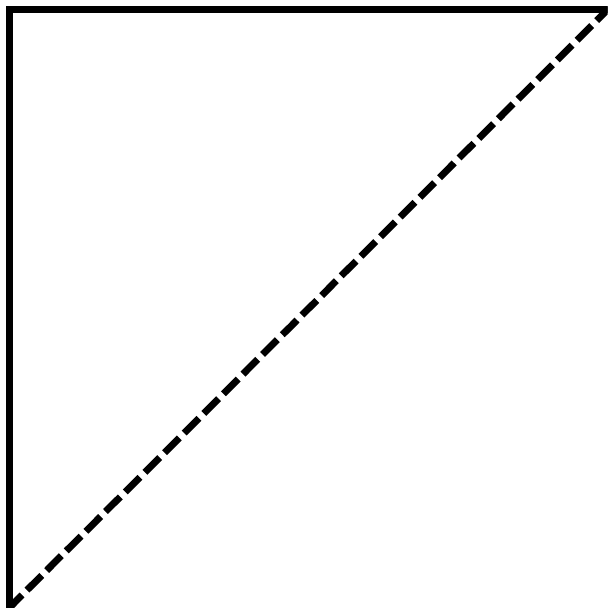}}\hspace{0.5 in}
\subfigure[]{\includegraphics[scale = .308]{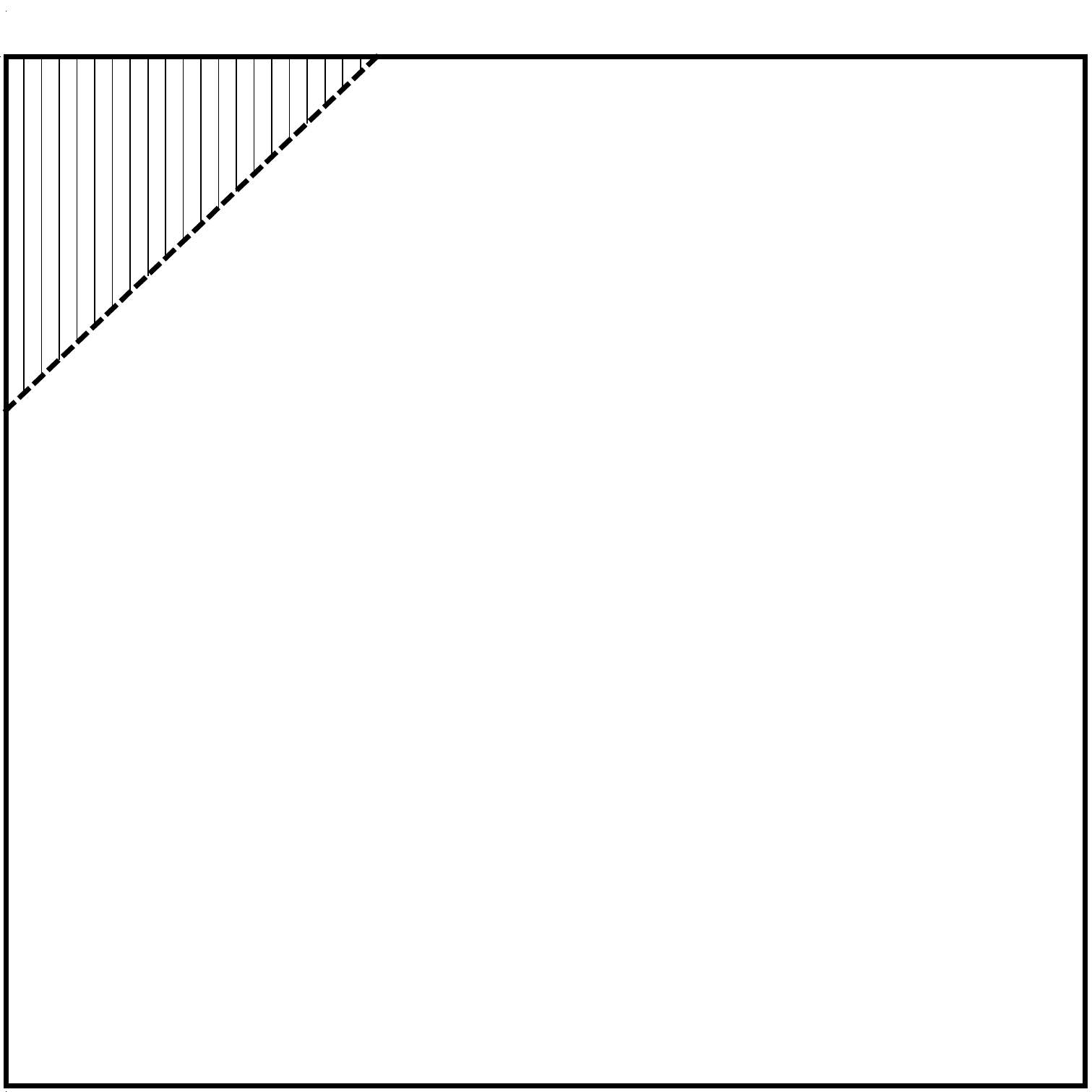}}
\end{center}
\caption{(a) The parent de~Sitter space with the future of the nucleation event removed and (b) the bubble universe are shown as separate conformal diagrams which are each portions of the Einstein static universe.  After an additional conformal transformation of the bubble universe (shaded triangle), the diagrams can be smoothly matched along the domain wall.  The resulting diagram (c) has a round $S^3$ as future infinity but is no longer a portion of the Einstein static universe.}
\end{figure}

The conformal diagram is left unchanged outside the future light-cone of the nucleation event, where the physical spacetime is also unchanged.  That is, we shall use the same conformal factor as before for this region.  

The future of the nucleation event is an open FRW universe, which either crunches (if $\Lambda<0$) or becomes asymptotically de~Sitter in the future (if $\Lambda>0$).  We show in Appendix~\ref{sec-transfrw} that this spacetime can be conformally mapped to a finite unphysical spacetime with the following properties:
\begin{itemize}
\item The conformal factor is smooth ($C^n$, with $n$ arbitrarily large) in the physical spacetime and continuous when extended to the future boundary.
\item The induced metric along the future light-cone of the nucleation event agrees with the induced metric on the same hypersurface in the outside portion of the diagram, which was left unchanged.
\item The future boundary is identical to the piece of the future boundary that was removed from the old diagram.
\end{itemize}
(Note that this portion of the unphysical spacetime will not be a portion of the Einstein static universe, nor is there any reason to demand that it should be.)

Because of the above properties, we are able to combine the remainder of the old Penrose diagram with the new portion covering the interior of the bubble.  This results in a single Penrose diagram which is smooth everywhere\footnote{There is enough freedom in the choice of the conformal factor in the bulk to ensure that the conformal factor is not only continuous but arbitrarily differentiable at the seam where the FRW bubble and the old de~Sitter parent are matched, while maintaining the round three-sphere metric of the future boundary.  Thick-wall bubbles and non-zero initial bubble radii can similarly be accommodated.} in the physical spacetime, and whose future boundary is a round three-sphere of unit radius.  Thus, it satisfies the condition we have imposed, that the Ricci scalar of the boundary metric be constant.

We can now simply iterate this procedure.  Complete the following sequence of operations, starting with $i=0$ (or really with $i=1$ since we have already gone through one iteration, above):
\begin{enumerate}
\item Using the conformal diagram you have constructed, with a future boundary metric satisfying $R=$ const, construct light-cone time for the $i$-bubble multiverse.
\item Identify the earliest bubble (i.e., the nucleation event with the smallest light-cone time) that has not already been included, and replace the future of the nucleation event with an FRW universe, which will be treated as completely stable in this step.  The resulting spacetime is the $i+1$-bubble multiverse.
\item Construct a new conformal diagram with constant Ricci scalar on the future boundary.  Since we have assumed that decay rates are small, decays will happen in asymptotically de~Sitter regions, where the construction given above for $i=0$ can be applied.
\item Increase $i$ by 1 and go back to step 1.
\end{enumerate}

Note that our construction is cumulative: every time a new bubble is included, the conformal factor is modified only inside this new bubble.  Moreover, the geometry at future infinity remains a unit round three-sphere throughout the process; the only effect of each step is that some part of future infinity is now in a different vacuum.  On the boundary, this allows us to think of bubble nucleation as the insertion of a disk of a corresponding size.  We will exploit this fact in the next section, when we describe eternal inflation and its attractor behavior from the boundary point of view.

\section{Properties of new light-cone time}
\label{sec-prop}

In this section, we will derive and solve the equation governing the distribution of vacua as a function of light-cone time, and we will find an expression for the number of events of type $A$ as a function of light-cone time.

\subsection{Rate equation}
\label{sec-rate}

We will work in the approximation of small decay rates, so that we can neglect bubble nucleations that happen during the early, non-vacuum-dominated phase of each FRW universe.  Moreover, we will neglect the small fraction of bubble nucleations that would lead to collisions with existing bubbles.  In the spirit of our new definition of light-cone time, however, we will express the rate equations in terms of variables that can be thought of as living on the future boundary: the volume $V_\alpha$ taken up by bubbles of type $\alpha$ on the boundary, and the variable $\eta=\exp(-t)$, which has a bulk interpretation as a time variable but also a boundary interpretation as a distance scale.

Consider a bubble of a de~Sitter vacuum $\alpha$ that nucleated at some late time $\eta_{\rm nuc}\ll 1$, so that its size on future infinity is $V_\alpha^{(1)}=\frac{4\pi}{3}\eta_{\rm nuc}^3$.  Now consider some later time $\eta/\eta_{\rm nuc}\ll 1$ inside the bubble.  The metric is approximately
\begin{equation}
ds^2 = {H_\alpha^{-2} \over \eta^2} \left( - d \eta^2 + d \vec x^2 \right)~,
\label{eq-asymet}
\end{equation}
where the comoving coordinate ${\mathbf x}$ ranges over a volume $V_\alpha^{(1)}$.  Here we have neglected the fact that at any finite $\eta$, the bubble will not have expanded to its full asymptotic comoving volume, and that the bubble is not exactly de~Sitter but may contain matter and radiation.  Both of these corrections decrease like powers of $\eta/\eta_{\rm nuc}$; in particular, most of the physical volume at small $\eta/\eta_{\rm nuc}$ will be empty de~Sitter space.  

The above metric is valid at zeroth order in the decay rate expansion: we have neglected all decays that might have already taken place in the bubble by the time $\eta$.  We will now include decays at first order.  The proper three-volume of the bubble at the time $\eta$ is $H_\alpha^{-3} V_\alpha^{(1)}/\eta^{3}$.  One Hubble volume occupies proper volume $ \frac{4\pi}{3}H_\alpha^{-3}$, so that the number of Hubble volumes is
\begin{equation}
n_\alpha^{(1)}=\frac{V_\alpha^{(1)}}{\frac{4\pi}{3} \eta^3}=\frac{\eta_{\rm nuc}^3}{\eta^3}~.
\label{eq-na1}
\end{equation} 
We note the relation $H_\alpha dt_{\rm prop}=dt=-d\eta/\eta$ between the proper time, $t_{\rm prop}$, light-cone time, $t$, and $\eta$.  We also recall the definition that $\kappa_{i\alpha}$ is the rate at which $i$-bubbles are nucleated inside the $\alpha$-vacuum, per Hubble volume and Hubble time.   In a single bubble of type $\alpha$, therefore, the total number of bubbles of type $i$ that are produced during a time $d\eta$ is 
\begin{equation}
dN_{i\alpha}^{(1)} = \kappa_{i\alpha} n_\alpha^{(1)} dt =  { -d\eta \over \frac{4\pi}{3} \eta^4}\,\kappa_{i\alpha} V_\alpha^{(1)} ~.
\label{eq-dnia1}
\end{equation}
The total number of bubbles of type $i$ produced inside $\alpha$-bubbles during the time $d\eta$ is obtained by summing over all $\alpha$-bubbles:
\begin{equation}
dN_{i\alpha} = {-d\eta \over \frac{4\pi}{3} \eta^4} \,\kappa_{i\alpha}V_\alpha(\eta) ~,
\label{eq-dnia}
\end{equation}
where $V_\alpha(\eta)$ is the total volume taken up at future infinity by $\alpha$ bubbles nucleated prior to $\eta$.

Now consider the total volume $dV_i^+$ taken up at future infinity by all the $i$-bubbles produced (in {\em any\/} vacuum) during the time $d\eta$.  Since we are neglecting bubble collisions, this is equal to the number of bubbles produced, times the volume taken up by each bubble:
\begin{equation}
dV_i^+= \frac{4\pi}{3} \eta^3 \sum_\alpha dN_{i\alpha}~.
\end{equation}
Though some vacua with $\Lambda\leq 0$ can decay, this effect is not important and will also be neglected.  Thus the sum runs only over de~Sitter vacua, whereas $i$ denotes a vacuum of any type.

As a result of decays into other vacua, the total boundary volume taken up by the de~Sitter vacuum $\alpha$ decreases during the time $d\eta$, by a volume $\frac{4\pi}{3} \eta^3$ per decay:
\begin{equation}
dV_\alpha^-=-\frac{4\pi}{3} \eta^3 \sum_i dN_{i\alpha}~.
\end{equation}
Focussing now on de~Sitter vacua alone, we can find that the total rate of change of the boundary volume occupied by $\alpha$-bubbles, $dV_\alpha=dV_\alpha^++dV_\alpha^-$, by combining the previous three equations:
\begin{equation}
dV_\alpha=\frac{-d\eta}{\eta} \sum _\beta M_{\alpha\beta} V_\beta~,
\label{eq-rb}
\end{equation}
where 
\begin{equation}
M_{\alpha\beta}\equiv \kappa_{\alpha\beta}-\kappa_\alpha\delta_{\alpha\beta}~.
\label{eq-mab}
\end{equation}

This is the rate equation for the volume occupied by metastable de~Sitter vacua {\em on the boundary\/}.  We derived this equation using bulk dynamics, which we understand at the semi-classical level.  But we have expressed it in terms of variables that are defined on the boundary: $V_\alpha$ is a boundary volume and $\eta$ is a boundary scale.  In this form, the rate equation can be naturally interpreted in terms of the boundary dynamics we seek.  In effect, we have derived a procedure for constructing the boundary fractal, working from the largest bubbles to the smallest.  Begin with the boundary sphere all in one ``color'', corresponding to the initial metastable de~Sitter vacuum in the bulk.  Now use the bubble distribution (\ref{eq-dnia1}) to include new bubbles, one by one, corresponding to disks of different colors. The disks are smaller and smaller (of radius $\eta$) as we progress to smaller scales $\eta$.  Just as there will be a finite number of bubbles prior to the time $t$ in the bulk, there will be a finite number of disks on the boundary for any UV cut-off $\eta$.  Their distribution will obey Eq.~(\ref{eq-rb}).

Eq.~(\ref{eq-dnia1}) specifies how many $i$-disks of a given size should be inserted in an $\alpha$-disk, but we also should specify {\em where\/} they should be inserted.  Since the nucleation rate is homogeneous in the bulk metric, Eq.~(\ref{eq-asymet}), new bubbles be inserted with equal probability in any infinitesimal volume occupied by the $\alpha$-disk.  We have excluded bubble collisions; this can be incorporated from the boundary point of view by forbidding the addition of any new disk whose boundary would overlap with an existing disk.  This amounts to excluding a zone of width $\eta$ neighboring all existing disk boundaries.  As we discussed in the previous section, the solution to the Yamabe problem on the sphere is not unique, and we still have the freedom to act with the group of global conformal transformations $SO(4,1)$.  However, this has no effect on Eq.~(\ref{eq-dnia1}), which is conformally invariant.

This completes our discussion of the boundary interpretation of the rate equation.  The boundary process we have described should emerge from a more general boundary theory that remains to be discovered, and one would expect Eq.~(\ref{eq-rb}) to constrain the construction of such a theory.  Meanwhile, in order to compare our result to the rate equation for ``old'' light-cone time, we would like to rewrite it using bulk variables.  In terms of light-cone time, Eq.~(\ref{eq-rb})  becomes
\begin{equation}
\frac{dV_\alpha}{dt}= \sum _\beta M_{\alpha\beta} V_\beta~.
\label{eq-mix}
\end{equation}
We can use Eq.~(\ref{eq-na1}) to eliminate $V_\alpha$, the volume taken up by $\alpha$-disks on the boundary with UV-cutoff $\eta=\exp(-t)$, in favor of the number of horizon volumes of type $\alpha$ in the bulk at the time $t$:
\begin{equation}
V_{\alpha}(t)=n_\alpha(t)\, \frac{4\pi}{3} \exp(-3t)~.
\label{eq-na}
\end{equation}
Thus, we obtain the bulk rate equation
\begin{equation}
\frac{dn_\alpha}{dt}=(3-\kappa_\alpha) n_\alpha+\sum _\beta \kappa_{\alpha\beta} n_\beta~.
\label{eq-nadot}
\end{equation}
This is identical to the rate equation for old light-cone time, Eq.~(36) in Ref.~\cite{BouYan09}.  Thus, at this level of approximation, the two definitions of light-cone time yield precisely the same rate equations.

\subsection{Attractor solution}
\label{sec-asol}

The solutions to the above rate equation exhibit attractor behavior at late times ($\eta\to 0$):\footnote{See Eq.~(37) of Ref.~\cite{GarSch05}, who analyzed its solutions.  Although our Eq.~(\ref{eq-mix}) takes the same mathematical form, it should be noted that it is for a different physical variable: the boundary volume fraction occupied by $\alpha$-disks, which by Eq.~(\ref{eq-na}) corresponds to a fraction of Hubble volumes in the bulk.  (In Ref.~\cite{GarSch05}, the relevant variable was the fraction of proper bulk volume occupied by vacuum $\alpha$.)}
\begin{equation}
V_\alpha(\eta)= \check V_\alpha \eta^q+O(\eta^{\bar q})~;
\label{eq-ar}
\end{equation}
or, in terms of bulk variables, 
\begin{equation}
n_\alpha(t) = \check n_\alpha e^{(3-q) t}+ O(e^{(3-\bar q) t})~.
\label{eq-h2}
\end{equation}
Here $-q$ is the largest eigenvalue of the matrix $M_{\alpha\beta}$, $\check V_\alpha$ is the corresponding eigenvector, and $\check n_\alpha\equiv V_\alpha/\frac{4\pi}{3}$; $-\bar q$ is the second-largest eigenvalue, and $0<q<\bar q$~\cite{GarSch05,BouYan09}.  

At short distances on the boundary (late times in the bulk), the distribution of disks (Hubble volumes) is governed by the ``dominant eigenvector'', $\check V_\alpha$, which can be thought of as the linear combination of de~Sitter vacua that decays most slowly.  In a realistic landscape, this eigenvector will have almost exclusive support in a single vacuum, which we shall denote by a star:
\begin{equation}
\check V_\alpha\approx v \delta_{\alpha*}~.
\end{equation}
This dominant vacuum is the slowest-decaying de~Sitter vacuum of the landscape~\cite{SchVil06}.  The normalization, $v$, of the eigenvector depends on the infrared boundary configuration (initial conditions in the bulk).

\subsection{Event counting}
\label{sec-count}

So far we have discussed only de~Sitter vacua; and even in asymptotically de~Sitter  bubbles, we have neglected any initial, transitory period during which the bubble universe might have been dominated by, say, matter, radiation, or slow-roll inflation.  This was sufficient for deriving the rate equation and the asymptotic attractor distribution for de~Sitter vacua.  As we now turn to the question of counting events in the multiverse, we must include all bubbles and regimes within them.  

Let $A$ be some type of event (e.g., a supernova occurs, or the microwave background temperature is found to be between $2.5\,$K and $3\,$K by some experiment).
Let $N_A(\eta)$ be the number of times an event of type $A$ has occurred in the bulk prior to the cut-off $\eta$.  (On the boundary, we expect that this is the number of times it has been encoded in modes of size greater than $\eta$.  For now, we will use the bulk definition since we know how to compute it.)  

To compute $N_A(\eta)$, we substitute the asymptotic distribution of de~Sitter vacua, Eq.~(\ref{eq-ar}), into our result for the number of bubbles of type $i$ produced per unit time $d\eta$, Eq.~(\ref{eq-dnia}):
\begin{equation}
dN_i= {-d\eta \over \frac{4\pi}{3} \eta^{4-q}} \sum_\alpha \kappa_{i\alpha} \check V_\alpha~;
\end{equation}
then we sum over all bubbles and nucleation times:
\begin{equation} 
  N_A(\eta) = 
  \sum_{i} \int_{\eta_0}^{\!\!\!\!\!\!\!\! i~\,\eta} 
  \left( \frac{dN_A}{dN_i}\right)_{\eta/\eta_n}
  \left( \frac{dN_i}{d\eta}\right)_{\eta_n}  d\eta_n ~,
  \label{eq-nie}
\end{equation}
where $dN_A/dN_i$ is the expected number of events of type $A$ that have occurred by the time $\eta$ in a single bubble of type $i$ nucleated at the time $\eta_n$.   The lower limit of integration, $\eta_0$, is an arbitrary early-time cut-off (an infrared cut-off on the boundary) that will drop out of the limit in Eq.~(\ref{eq-p}).

At late times, all bubbles of type $i$ are statistically equivalent, because they are produced locally in an empty de~Sitter region.  Therefore, $dN_A/dN_i$ depends only on the ratio
\begin{equation}
\zeta\equiv \eta/\eta_n~.
\end{equation}
To avoid overcounting, the integral should run only over a single bubble of vacuum $i$, excluding regions of other vacua nucleated inside the $i$ bubble; this restriction is denoted by index $i$ appearing on the upper left of the integration symbol. 

Combining the above results, and changing the integration variable from $\eta_n$ to $\zeta$,
\begin{equation}
N_A(\eta) = \frac{1}{\frac{4\pi}{3} \eta^{3-q}} \sum_i \sum_\alpha N_{Ai} \kappa_{i\alpha}\check V_\alpha~.
\label{eq-nie2}
\end{equation}
Here we have assumed for simplicity that $A$ cannot occur in the dominant vacuum, $*$ (otherwise, see Ref.~\cite{BouYan09}); and we have defined
\begin{equation}
  N_{Ai}\equiv
  \int_0^{\!\!\!\!\!\!\!\! i~\,1}\left( \frac{dN_A}{dN_i}\right)_\zeta \zeta^{2-q} d\zeta ~.
\label{eq-nim}
\end{equation}
The lower limit of integration in Eq.~(\ref{eq-nim}) should strictly be $\zeta=\eta/\eta_0$, so this result is valid only at late times ($\eta\to 0$), but this is the only regime relevant for computing relative probabilities.

The quantity $\left( \frac{dN_A}{dN_i}\right)_\zeta$ depends on the details of the bubble universe $i$ and on how the cut-off surface $\eta$ intersects this bubble.  
It can be written as
\begin{equation}
\left( \frac{dN_A}{dN_i}\right)_\zeta=\int_\zeta^{\!\!\!\!\!\!\!\! i~\,1} d\zeta' \,d^3 {\mathbf x}\, \sqrt{g}\, \rho_{A i}(\zeta', {\mathbf x})~.
\label{eq-niz}
\end{equation}
where $\rho_{A i}$ be the density of events of type $A$ per unit four-volume in a bubble $i$, and $g$ is the determinant of the metric.  The integral is over the interior of a single bubble of type $i$, from its nucleation up to a light-cone time a factor $\zeta$ after nucleation.  Substituting into Eq.~(\ref{eq-nim}) and exchanging the order of integration, we obtain an alternative expression for $N_{A i}$:
\begin{eqnarray} 
N_{A i} & = & \int_0^{\!\!\!\!\!\!\!\! i~\,1} d\zeta' \int d^3 {\mathbf x}  \,\sqrt{g} 
\,\rho_{A i}(\zeta', {\mathbf x} ) \int_{0}^{\zeta'} d\zeta \,\zeta^{2-q} 
\label{eq-nim1}\\
& = & \frac{1}{3-q}\int d^4x \, \sqrt{g}\, \rho_{A i}(x) \,\zeta(x)^{3-q}~,
\label{eq-nim2}
\end{eqnarray} 
where the integral is over the entire four-volume of a single $i$-bubble, with the exclusion of new bubbles nucleated inside it.

\section{The probability measure}
\label{sec-pm}

In Sec.~4.1 we derive the probability measure defined by the new light-cone time cut-off.  In the following subsections, we show that it is equivalent to probabilities computed from three other measure prescriptions (see Table~\ref{square}).  

\subsection{Probabilities from the new light-cone time cut-off}

The relative probability of two events $A$ and $B$ is defined by
\begin{equation}
\frac{p_A}{p_B}=\lim_{\eta\to 0}\frac{N_A(\eta)}{N_B(\eta)}~.
\label{eq-pp}
\end{equation}
By substituting Eq.~(\ref{eq-nie2}), we find that the probability of an event of type $A$ is given by
\begin{equation}
p_A\propto \check N_A \equiv 
\sum_i \sum_\alpha N_{Ai} \kappa_{i\alpha}\check V_\alpha~.
\label{eq-pp2}
\end{equation}
[We recall here that $\check V_\alpha$ is the eigenvector with largest eigenvalue of the matrix $M_{\alpha\beta}$ given in Eq.~(\ref{eq-mab}); $\kappa_{i\alpha}$ is the dimensionless nucleation rate of $i$-bubbles in the de~Sitter vacuum $\alpha$, and $N_{A i}$ can be computed from Eqs.~(\ref{eq-nim}) or (\ref{eq-nim2}); below we shall discover a simpler way of computing it, from Eqs.~(\ref{eq-nim3}) and (\ref{eq-vcp}).]  This expression is not normalized, but ratios can be taken to obtain relative probabilities.

We should go on to compute some probability distributions of interest.  We should verify that catastrophic predictions such as Boltzmann brains, the youngness paradox (``Boltzmann babies''), and the $Q$-catastrophe are absent.  If so, we should then move on to compute probability distributions over certain quantities of interest, such as the cosmological constant, the primordial density contrast, the spatial curvature, and the dark matter fraction, and we could verify that observed values are not unlikely under these distributions (say, in the central $2 \sigma$).  Instead, we will now demonstrate that the new light-cone time cut-off is equivalent to other measures that have already been studied.

\begin{table}[t] 
\centering 
\begin{tabular}{l c c c c} 
& use boundary &  & use geodesics \\[0.5ex]	
\hline 
Global measures:~~~ &  New Light Cone Time &  $\Leftrightarrow$ & Old Light Cone Time \\
& $\Updownarrow$ & $\Nwsearrow$ & \\ 
Local measures:~~~ & New Causal Patch &  & Old Causal Patch \\ \hline \end{tabular} \label{square} 
\caption{Equivalences between measures.  The new light-cone time cut-off is equivalent to probabilities computed from three other measure prescriptions (double arrows).  This implies that all four measures shown on this table are equivalent in the approximation described at the beginning of Sec.~2.3.  (See Ref.~\cite{BouYan09} for a more general proof of the equivalence between the old light-cone time cut-off and the old causal patch measure.)}
\end{table}

\subsection{Equivalence to the old light-cone time cut-off}
\label{sec-gloglo}

Probabilities computed from the old light-cone time cut-off are also of the general form (\ref{eq-pp2})~\cite{BouYan09}.  We will now show that they are in fact identical, under the assumptions made in the previous sections. 

We consider each factor appearing inside the sum in Eq.~(\ref{eq-pp2}) in turn.  The decay rate $\kappa_{i\alpha}$ is a property of the landscape vacua unrelated to the definition of light-cone time.  Moreover, we have shown that new light-cone time gives rise to the same rate equation and attractor solution as old light-cone time~\cite{Bou09,BouYan09}, assuming that all metastable de~Sitter vacua are long-lived.   This implies that the eigenvector $\check V_\alpha$ is the same, whether old or new light-cone time is used.  Therefore any difference between the two measures would have to come from the quantity $N_{A i}$.  However, if we approximate bubble interiors as homogeneous, isotropic, open universes, with metric of the form
\begin{equation}
ds^2=-d\tau^2+a(\tau)^2 (d\chi^2+\sinh^2\chi\, d\Omega_2^2)~.
\end{equation}
then no such difference arises.  At this level of approximation, old and new light-cone time define the same probability measure.



To demonstrate this, let us rewrite $N_{A i}$, starting from Eq.~(\ref{eq-nim2}).
Homogeneity implies that $\rho_{A i}$ depends only on the FRW time, $\tau$.  Therefore, we can write
\begin{equation}
N_{A i}=\frac{1}{3-q}\int_0^\infty d\tau\, \sigma_{A i}(\tau)\, {\cal V}_c(\tau)~,
\label{eq-nim3}
\end{equation}
where $\sigma_{A i}(\tau)\equiv \rho_{A i}(\tau) a(\tau)^3$ is the density of events of type $A$ per FRW time and per unit comoving volume, and
\begin{equation}
{\cal V}_c(\tau)\equiv \int
d\tilde\Omega_3~ \zeta(\tau,\chi)^{3-q}~.
\label{eq-vct}
\end{equation}
Stricly, the integral should exclude comoving regions that have already decayed.  But since the decay probability is homogeneous, we may absorb these excisions into a decrease of the comoving density $\sigma_{A i}$ and let the integral run over the entire unit hyperboloid, $d\tilde\Omega_3= 4\pi \sinh^2\chi \,d\chi$.  The integral converges because the factor $\zeta^{3-q}$ rapidly vanishes at large $\chi$.  Thus, ${\cal V}_c(\tau)$ can be thought of as an effective comoving volume whose events contribute to $N_{A i}$ at the time $\tau$.

Since $q$ is the decay rate of the longest-lived vacuum, we can assume that it is exponentially small in a realistic landscape, so it can be neglected in Eq.~(\ref{eq-vct}).  Moreover, from the definition of light-cone time in terms of volumes $V$ on the future boundary, it follows that 
\begin{equation}
\zeta^3=e^{-(t-t_n)}=V(t)/V(t_n)~.
\label{eq-zev}
\end{equation}
Note that these equalities can be taken to hold by definition, {\em no matter how the volume on the future boundary is defined.}  We will now abandon the particular definition made in Sec.~\ref{sec-ricci} and consider a general metric $G_{\mu\nu}(\mathbf{y})$ on the future boundary of the multiverse.  We shall find that ${\cal V}_c$, and thus $N_{A i}$, is independent of this choice of metric if the bubble is homogeneous.

Upon substitution of Eq.~(\ref{eq-zev}), the integrand in Eq.~(\ref{eq-vct}) is the boundary volume of the future light-cone $L(\tau,\chi)$ of each point, which we may in turn write as an integral over the region of the boundary enclosed by the light-cone:
\begin{eqnarray}
{\cal V}_c(\tau) & = & V(t_n)^{-1}\int d\tilde\Omega_3 \int_{L(\tau,\chi)} d^3y \sqrt{G}\\
& = & V(t_n)^{-1} \int_{L(0,\chi)} d^3y \sqrt{G} \int_{CP(y,\tau)} d\tilde\Omega_3 ~.
\label{eq-lkj}
\end{eqnarray}
In the second line, we exchanged the order of integration.  The ranges of integration are the crucial point here.  The union of the future light-cones from all points on the hyperboloid covers precisely the entire disk corresponding to the bubble universe at future infinity.  Hence, the outer integral ranges over this disk, which we have written as the region enclosed by the future light-cone of the bubble nucleation event, $L(0,\chi)$.  But each boundary point is covered by many future light-cones of bulk points.  Correspondingly, the inner integral ranges over all points on the FRW slice whose future light-cone includes the boundary point $y$.  But this is simply the set of points in the intersection of the $\tau=$ const.\ slice with the causal past of $y$, i.e., the portion of the constant FRW time slice that lies in the causal patch whose tip is at $y$.  By homogeneity, its volume is independent of $y$ and will be denoted $V^{\rm CP}_c(\tau)$.  Thus, the two integrals factorize.  Since $\int_{L(0,\chi)} d^3y \sqrt{G}=V(t_n)$ by definition, we obtain
\begin{equation}
{\cal V}_c(\tau)=V^{\rm CP}_c(\tau) ~.
\label{eq-vcp}
\end{equation}
With this result, Eq.~(\ref{eq-nim3}) becomes manifestly independent of the definition of boundary volume.  Thus, the quantity $N_{A i}$ will be the same for any type of ``light-cone time'', if the bubbles of type $i$ are homogeneous FRW universes.  

This is a remarkably general result, so let us state it very clearly.  From Eq.~(\ref{eq-nim}), $N_{A i}$ would appear to depend on the definition of $\zeta$, which in turn depends on the definition of the scale $\eta$ associated with a bulk point $E$ as $V(E) = \frac{4\pi}{3} \eta^3$, where $V$ is the volume taken up by the causal future of $E$ on the future boundary.  This volume, of course, can be defined in different ways: for example, in Ref.~\cite{Bou09} it was defined by projecting onto a fiducial bulk hypersurface, whereas in the present paper it was defined in terms of the induced metric of the boundary in a conformal frame with $R=$ const.  These different definitions of $V$ generically do lead to different values of $\zeta$, $\eta$, and light-cone time $t$ at the event $E$.  What we have shown is that they cannot lead to any difference in $N_{A i}$.  

In the homogeneous FRW approximation, then, different definitions of light-cone time can only lead to a different probability measure if they lead to different rate equations.  This can certainly happen in principle.  However, we have shown earlier that the two definitions given so far---``old'' and ``new'' light-cone time---yield the same rate equations.  We conclude that they also lead to the same probability measure, if bubble universes are approximated as homogeneous open FRW universes.

\subsection{Equivalence to the new causal patch measure}
\label{sec-gloloc}

Consider an ensemble of causal patches whose tips are distributed over the future boundary of the multiverse, at constant density $\delta$ with respect to the boundary metric defined in Sec.~\ref{sec-ricci}.  The probability for an event of type $A$, according to the causal patch measure, is proportional to the ensemble average of the number of times an event of type $A$ occurs inside a causal patch, $\langle N_A \rangle^{\rm CP}$.  Let $\eta_0$ be an early-time cut-off.  Then
\begin{equation}
p_A^{\rm CP}\propto \langle N_A \rangle^{\rm CP}\propto \int_0^{\eta_0} d\eta\, \frac{dN_A}{d\eta}\, \eta^3~.
\end{equation}
The quantity $d\eta \frac{dN_A}{d\eta}$ is the number of events that happen during the interval $d\eta$.  Each of these events in contained in $\frac{4\pi}{3} \eta^3 \delta$ causal diamonds, since $\frac{4\pi}{3} \eta^3$ is the volume of the causal future of an event at time $\eta$, and the causal diamonds that contain a given event are precisely those which have tips in the events future~\cite{BouYan09}.

Suppose that $\eta_0$ is small enough to lie deep inside the asymptotic attractor regime, where Eq.~(\ref{eq-nie2}) holds.  Then the above integral can be evaluated, and we find
\begin{equation}
p_A^{\rm CP}\propto \check N_A~.
\label{eq-pnewcp}
\end{equation}
Comparison with Eq.~(\ref{eq-pp2}) reveals that the causal patch measure gives the same relative probabilities as the new light-cone time cut-off, if initial conditions for the causal patch are chosen in the attractor regime.  

This result is similar to the global-local equivalence proven for the old light-cone time cut-off~\cite{BouYan09}.  The local dual in that case was also an ensemble of causal patches beginning on a bulk hypersurface with vacuum distribution in the attractor regime, Eq.~(\ref{eq-h2}).  However, the ensemble itself was selected by erecting geodesics at fixed density per unit Hubble volume on the initial hypersurface.   By contrast, the causal patch ensemble we derived above from the new light-cone time is defined in terms of a uniform distribution of {\em tips\/} of causal patches per unit boundary volume, not of the starting points of their generating geodesics on a bulk hypersurface.  

In fact, the concept of a generating geodesic of a causal patch appears to be unnecessary in the ``new'' causal patch measure.  Instead, the patches are defined more simply as the past of points on the future boundary, or TIPS~\cite{GerPen72}.  This simplification is the local analogue of the elimination, in our definition of new light-cone time, of the family of geodesics that was needed to define old light-cone time.  We regard these simplifactions as a significant formal advantage of the boundary viewpoint.

\subsection{Equivalence to the old causal patch measure}
\label{sec-gloold}

We turn to the diagonal arrow in the duality square of Table~\ref{square}, the equivalence of the new light-cone time cut-off to the ``old'' causal patch measure, defined by erecting a family of geodesics orthogonal to a late time attractor hypersurface and constructing the causal patch of each geodesic, i.e., the past of its endpoint.  This equivalence follows by combining two other arrows:  the equivalence of the old and new light-cone time cut-offs (Sec.~\ref{sec-gloglo}), and that of the the old light-cone time cut-off and the the (old) causal patch measure~\cite{BouYan09}.  (This latter proof, unlike any of the arguments in this section, requires no simplifying assumptions such as homogeneity of the bubbles.) 
In the interest of a self-contained presentation, we will now present a shortcut that directly establishes the equivalence.

A geodesic starting in a vacuum $\alpha$ with probability $\check V_\alpha$ will enter vacuum $i$ an expected number $\sum_\alpha \kappa_{i\alpha} \check V_\alpha$ of times~\cite{BouYan07}.  (Since vacua are unlikely to be entered twice along the same geodesic, this expectation value can be thought of as an unnormalized probability of entering $i$.)  By the assumed homogeneity, all causal patches with tips in the same FRW bubble have statistically identical content, which, by Eq.~(\ref{eq-nim3}), is given by $(3-q) N_{A i}$.  Thus, the probability of an event of type $A$, according to the the ``old'' causal patch measure, is
\begin{equation}
p_A^{\rm old\,CP}\propto \sum_i \sum_\alpha N_{Ai} \kappa_{i\alpha}\check V_\alpha~,
\label{eq-poldcp}
\end{equation}
which agrees with the probability computed from the new light-cone time cut-off, Eq.~(\ref{eq-pp2}).

\section{The general case}
\label{sec-general}

At first sight, it seems that our prescription is well-defined for any bulk spacetime. Simply find a conformal transformation which makes the bulk spacetime finite, and then use the constant scalar curvature condition to fix the ambiguity in the boundary metric. The Yamabe theorem would seem to guarantee that this can always be done. But this process is not as simple as it sounds once we go beyond the approximation of homogeneous FRW universes stated in Sec.~\ref{sec-frw}.  

In this section, we will discuss the additional challenges that arise in the general case, how they might be addressed, and what they may imply for the phenomenology of new light-cone time.  We first discuss perturbative inhomogeneities which complicate the story but do not obviously lead to a pathology. Then we discuss situations where the boundary metric is singular. These situations require a generalization of our prescription which we outline, but whose details we leave for the future.  Finally, we argue that the probabilities derived from new and old light-cone time will differ in the presence of inhomogeneities, with new light-cone time being favored phenomenologically.

\subsection{Perturbative inhomogeneities}

No bubble universe will be exactly homogeneous and isotropic.  Gravitational waves will perturb the metric and break spherical symmetry. Modes will leave the horizon during slow roll inflation and become frozen. We expect that such perturbations will affect the future boundary. Indeed, any metric of the form
\begin{equation}
ds^2 = -dt^2 + e^{2 H t} a_{ij}(x^k) dx^i dx^j~,
\label{eq-bougib}
\end{equation}
is a solution to Einstein's equation with positive cosmological constant in the limit $t \to \infty$, where $a_{ij}$ is an {\em arbitrary\/} 3-metric that depends only on the spatial coordinates $x^k$.\footnote{There are global constraints at finite time that may restrict the form of the boundary metric.  Though it seems to us implausible, we cannot rule out that such restrictions may prevent the appearance of smaller and smaller geometric features on the boundary.}  (This result is attributed to Starobinski in Ref.~\cite{BouGib82}.)  In terms of conformal time, the metric is
\begin{equation}
ds^2 = {1 \over \eta^2} \left( -d \eta^2 + a_{ij} dx^i dx^j
\right)~,
\end{equation}
and the future boundary is the surface $\eta=0$.  Rescaling by the
conformal factor $\Omega=\eta$, we obtain a (preliminary) boundary metric
\begin{equation}
ds^2 = a_{ij} dx^i dx^j~.
\end{equation}
Recall that $a_{ij}$ is arbitrary, so any metric is possible on future infinity. If Eq.~(\ref{eq-bougib}) describes the late time limit of a de~Sitter bubble universe, $a_{ij}$ will need to be further conformally rescaled to finite volume and matched to the outside portion of the boundary.  The condition $R=$ const.\ can then be achieved by a third conformal transformation.  But because not every metric $a_{ij}$ is conformal to a portion of the round $S_3$, in general the future boundary will not be a round $S_3$, and its geometry will contain features on the scale of the future light-cone of any perturbation.

Bubble nucleations occur at arbitrarily late times, corresponding to small sizes on the boundary. As a result, the boundary metric is perturbed on arbitrarily small scales.  The boundary {\em geometry\/} is a fractal.  It is important to understand the distinction we are drawing here to the homogeneous case.  In this case, too, a fractal develops on the future boundary, but that fractal describes the disks corresponding to different vacua distributed over a fixed background metric (a round $S^3$).  This is similar to the kind of fractal we are familiar with.  What we are finding in the general case is that in addition, the geometry itself becomes a fractal.  It exhibits self-similar features such as curvature on arbitrarily small scales.

By itself, the fractal curvature presents no serious obstacle.  It just forces us to construct the background geometry of the boundary the same way we constructed its ``matter content'' (say, the disks representing different vacua): by starting in the infrared and refining it step by step as we move to short distances.

However, there are additional problems.  Suppose we would like to include some bulk perturbation, such as the formation of a new bubble universe, on the boundary.  In the simplified multiverse we studied in Sec.~\ref{sec-frw}, the conformal transformation that achieved constant Ricci scalar acted nontrivially only in the future of the nucleation event; it did not change the conformal transformation outside the future light-cone.  But the Yamabe problem cannot generally be solved subject to such strong boundary conditions.  Thus, we expect that for general bulk perturbations, the boundary metric with $R=$ constant will be modified in regions that are outside the lightcone of the perturbation. So the nucleation of an inhomogeneous bubble generically changes the lightcone time of every event, even events out of causal contact with the nucleation.  

In the homogeneous approximation, if we are given some finite region of the bulk spacetime, we have enough information to compute the surfaces of constant light-cone time in that region because adding bubbles outside the region, even in its future, will not change the size of future lightcones.  If we had a computer simulating the bulk evolution, we could write a simple algorithm for generating the entire bulk spacetime underneath the cutoff surface.  But in the inhomogeneous case, we do not have a simple prescription for the size of a bulk event on the boundary. Events in the future, or even in causally disconnected regions, will change the boundary metric, and thus could change the size of ``old'' future light-cones on the boundary.  So given a computer which can simulate the bulk evolution, we do not have an algorithm for computing the bulk spacetime up to the cutoff, because it is unclear which events will turn out to occur at times after the cutoff.
 
This poses a challenge to the iterative construction of light-cone time, in Sec.~\ref{sec-frw}.  This procedure relied on moving forward along an existing light-cone time foliation, the $i$-bubble multiverse, and including the first new bubble nucleation reached in this manner to compute the $i+1$-bubble multiverse.  In this algorithm, the light-cone time was never modified for events prior to the new bubble nucleation.  Thus, we could be sure at each stage that we have a final version of the foliation up to the $i+1$-th bubble nucleation. As a result, a UV cut-off on the boundary had a simple bulk translation: bubble nucleations after a certain time should be ignored.   

In fact, it is unclear whether a discrete iterative process remains a viable approximation scheme in the general case.  We were able to focus on bubble nucleation events only, since by our assumptions no other events had the potential to modify the future boundary.  In general, one expects that all spacetime events, or at least all events in regions that are not empty de~Sitter space, will affect the future boundary.  This is a continuous set.  

While it is possible that some or all of the above effects are quantitatively negligible, they raise interesting conceptual issues.  If we had a boundary description of the theory, presumably it would be well defined to go down to shorter and shorter scales. What is missing at this point is a simple bulk prescription once we go beyond the approximation of homogeneous FRW universes. The fact that the boundary geometry becomes fractal once one includes all of the expected bulk dynamics means that defining volumes on the boundary is trickier. What is needed is a way to construct the UV-cutoff version of the boundary geometry, which is not a fractal. This can perhaps be done by beginning with the homogeneous approximation and then perturbing around it, but we have not attempted to do this.

\subsection{Singularities in the boundary metric}

\begin{figure}\label{fig-hatmultiverse}
\begin{center}
\includegraphics[scale = .35]{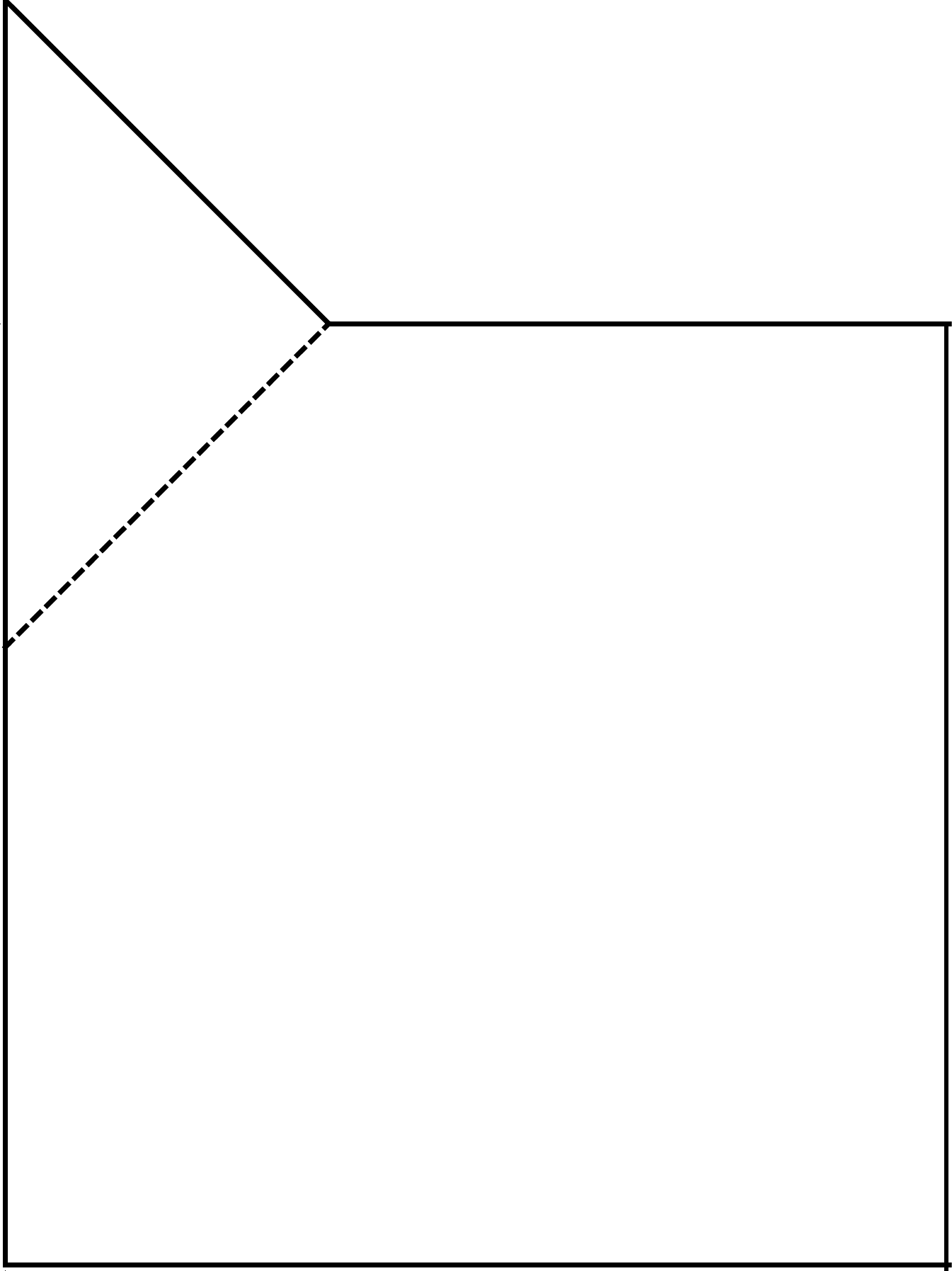}
\end{center}
\caption{Conformal diagram of de~Sitter space containing a bubble universe with $\Lambda=0$}
\end{figure} 
 There is a more severe pathology that can arise in constructing the boundary metric. What if the bulk spacetime does not lead to a boundary metric which is nonsingular and $3+1$ dimensional everywhere? There are several ways this can happen. Perhaps the simplest is the nucleation of a $\Lambda = 0$ bubble. Future infinity will now contain a ``hat'' as shown in Fig.~\ref{fig-hatmultiverse}. The induced metric will be degenerate because the ``hat'' is a null cone. The null portion of the boundary metric is
\begin{equation}
ds^2 = r^2 d\Omega_2^2 ~,
\end{equation}
which is not 3-dimensional. So a bulk spacetime that includes $\Lambda =0 $ asymptotic regions will have a singular boundary metric. It is then unclear how to define volumes on the boundary.

Another example of a bulk spacetime that leads to a singular boundary
metric is a Schwarzschild black hole in de Sitter space. The metric
for Schwarzschild-de Sitter in static coordinates is
\begin{equation}
ds^2 = -\left(1-\frac{2m}{r}-\frac{r^2}{l^2}\right)dt^2 + \frac{1}{1-\frac{2m}{r}-\frac{r^2}{l^2}} dr^2 +r^2  d\Omega_2^2~. 
\end{equation}
If we continue the metric behind the cosmological horizon and let
$r\rightarrow \infty$, the metric approaches pure de Sitter space.
 For large $r$, the metric is
\begin{equation}
ds^2 = - {l^2  \over r^2} dr^2  + r^2 \left({1 \over l^2} dt^2 +
d\Omega_2^2 \right)~.
\end{equation}
The boundary is $r \to \infty$; conformally rescaling the factor of
$r^2$ gives the boundary metric
\begin{equation}
ds^2 = {1 \over l^2} dt^2 +
d\Omega_2^2~.
\end{equation}
The $r= \infty$ part of the future boundary is a spatial cylinder. This is conformal to a sphere, aside from two missing points at the poles.

What about the other part of future infinity, the black hole
singularity? Here we take the opposite limit $r \to 0$ to obtain the
metric
\begin{equation}
ds^2 = -{r \over 2 m} dr^2 + {2 m \over r} \left(dt^2 + {r^3 \over 2 m}
d\Omega_2^2 \right)~.
\end{equation}
As $r \to 0$, the $t$ direction blows up while the $S_2$ becomes
small. Therefore, if we do the conventional conformal rescaling to
keep the $t$ direcion finite, the $S_2$ will have zero size on the
boundary, and the boundary metric will be one-dimensional,
\begin{equation}
ds^2 = dt^2~.
\end{equation}
So the boundary metric for a Schwarzschild-de Sitter black hole can be thought of as a sphere with a string coming out from the north pole. This is not a 3-dimensional manifold and it is not clear how to generalize our construction to this case.

Bubbles with negative cosmological constant end in a spacelike singularity.  In the homogeneous approximation these big crunches actually do not deform the boundary away from the round $S_3$. However, we know that the homogeneous approximation breaks down at late times. Instead of a homogeneous collapse, any small region near the singularity will be of the Kasner form, with two contracting and one expanding direction, much like the singularity of the Schwarzschild black hole discussed above. The boundary metric will be one-dimensional.  The Kasner behavior is uncorrelated between different parts of the singularity: which direction is expanding changes from place to place.  Therefore we expect that the boundary geometry for a big crunch is something like a fishnet.
 
One might expect that the light-cone time cut-off (and by extension, the causal patch) is simply inapplicable to portions of the multiverse that are in the domain of dependence of future singularities or hats~\cite{Bou09}, since the analogy with AdS/CFT breaks down entirely in such regions.  In this case, the measure may still give a good approximation to the correct probabilities in regions such as ours, which are in the eternally inflating domain (roughly, the de~Sitter regions).  Assuming that a smoothing procedure can be found along the lines of the previous subsection, one can ignore singularities while they are smaller than the UV cut-off and, if one accepts that the measure is meant to apply only to eternally inflating regions, one can stay away from the singularities once they are larger than the UV cut-off.

However, there is another way in which the boundary can have portions of different dimensionality.  In a realistic landscape, one expects that there will be de~Sitter vacua with dimension smaller or greater than $3+1$.  For example, in a landscape that includes both 4-dimensional and 6-dimensional de Sitter vacua, some regions of the future boundary will be 3-dimensional while others will be 5-dimensional.  We would need some way of generalizing the construction to boundaries with a variety of dimensionalities.

\subsection{Inhomogeneities}

From the previous subsections we conclude that two things are needed to make our prescription well-defined in a realistic multiverse. We must be able to allow the boundary to change dimensionality from place to place, and we need a procedure for smoothing the fractal nature of the boundary geometry.  We will now assume that this can be achieved and turn to the phenomenological implications of giving up the assumption of homogeneity.  

In Sec. ~\ref{sec-gloglo} we established the equivalence of old and new light-cone time in the homogeneous approximation.  It is precisely when homogeneity is broken that old and new light-cone time differ.  We will find it convenient to analyze this difference in the local version of each measure, the old and new causal patch measures.  A causal patch is always the same type of region: the past of a point on the future boundary.  But the old and new causal patch measure consider different ensembles of causal patches, and probabilities are computed as an ensemble average.  

In the old causal patch measure, the ensemble is defined by erecting timelike geodesics orthogonal to an initial hypersurface in the dominant vacuum, and following their evolution until they reach a crunch. The resulting patches will contain different decoherent histories. If a geodesic enters a bubble universe like ours, it quickly becomes comoving~\cite{BouFre07}.  With probability presumably of order one, it will become bound to a galaxy or galaxy cluster.  After the cosmological constant begins to dominate, at the time $\tau_\Lambda$, it may well remain bound to this structure for a time much greater than $\tau_\Lambda$, until it is either ejected by some process or falls into a black hole, or the structure disintegrates.  As a consequence, it may be the case that typical causal patches in the old ensemble do {\em not\/} become empty at late times but contain a single bound structure or black hole~\cite{PhiAlb09}. This could have important consequences for some probability distributions.  

For example, consider a class of observers that live at a FRW time $\tau_{\rm obs}$.  (Nothing else is assumed, e.g., one need not assume that they live in a vacuum like ours or are in any way similar to us~\cite{BouHar10}.)   What value of the cosmological constant are such observers likely to find?  The prior probability distribution in the landscape favors large $\Lambda$.   But for sufficiently large $\Lambda$, one has $\tau_\Lambda\ll \tau_{\rm obs}$.  In the homogeneous approximation, the number of observers inside the causal patch is then diluted by the de~Sitter expansion as $\exp(- \sqrt{3 \Lambda} \tau_{\rm obs})$, so that the probability distribution peaks at $\Lambda\sim \tau_{\rm obs}^{-2}$.  Thus the causal patch predicts that observers are likely to find themselves at the onset of vacuum domination~\cite{BouHar07}.  This predictions is both more general and more successful than the original prediction that $\Lambda\sim \tau_{\rm gal} ^{-2}$~\cite{Wei87}, where $\tau_{\rm gal}$ is the time of virialization; and it is more robust to the variation of other parameters such as the primordial density contrast.

In an inhomogeneous universe, however, the exponential decay is cut off when all but structures to which the generating geodesic was {\em not\/} bound have been expelled from the horizon.  After this time, the mass inside the causal patch does not decrease rapidly.  Of course, this still corresponds to a reduction in mass by a factor $10^{-11}$ in a universe like ours.  It will not spoil the prediction unless the number of observers per galaxy during the era $\tau\gg \tau_\Lambda$ is larger than in the present era by the inverse factor, which may well fail to be the case for observers like us in a universe like ours.  

But let us view the landscape as a whole and treat observers more abstractly.  The effect of inhomogeneity on the old causal patch ensemble is to remove the pressure for observations to take place not too longer after $\tau_\Lambda$.  This  allows for a much more efficient use of free energy, by conversion into quanta with wavelength as large as the de~Sitter horizon. If we model observers by entropy production as proposed in~\cite{Bou06,BouHar07}, and if we assume that processes that make such efficient use of free energy do not have small prior probability, then we should be rather surprised that we are not among this class of observers~\cite{PhiAlb09}.  

The new causal patch measure appears to resolve this puzzle.  Inhomogeneities such as galaxies and the black holes they contain will decay on a timescale that is power-law and thus negligible compared to typical vacuum decay timescales~\cite{Dys79}.  We expect, therefore, that they will leave little imprint on the future boundary of the multiverse, in the sense that the metric will not have strong features that favor causal patches containing structure at late times.  The fraction of causal patches containing any structure at times $\tau \gg \tau_\Lambda$ will be 
$\exp(- \sqrt{3 \Lambda} \tau_{\rm obs})$, so the above analysis, which was valid only  in the homogeneous approximation for the old causal patch measure, will always be valid in the new causal patch measure after ensemble-averaging.

\acknowledgments We would like to thank David Berenstein, Petr Ho\v{r}ava, Gary Horowitz, Shamit Kachru, Steve Shenker, Eva Silverstein, Lenny Susskind, and Edward Witten for discussions.  This work was supported by the Berkeley Center for Theoretical Physics, by the National Science Foundation (award number 0855653), by the Institute for the Physics and Mathematics of the Universe, by fqxi grant RFP2-08-06, and by the US Department of Energy under Contract DE-AC02-05CH11231.

\appendix

\section{Conformal factor for a homogeneous bubble universe}
\label{sec-transfrw}

In this appendix we show how to find a conformal transformation which maps a universe with a bubble nucleation event into a finite unphysical spacetime with a round $S^3$ as future infinity.
Our procedure is to begin with the full Penrose diagram of the parent de Sitter space, and then remove the portion which lies within the future light-cone of the nucleation event.
The removed piece will be replaced with a new unphysical spacetime which is conformally equivalent to an open FRW universe with nonzero cosmological constant and has the following properties, outlined in Sec.~\ref{sec-frw}:

\begin{itemize}
\item The conformal factor is smooth ($C^n$ with $n$ arbitrarily large) in the physical spacetime and continuous when extended to the future boundary.
\item The induced metric along the future light-cone of the nucleation event agrees with the induced metric on the same hypersurface in the outside portion of the diagram, which was left unchanged.
\item The future boundary is identical to the piece of the future boundary that was removed from the old diagram.
\end{itemize}
(Note that the new portion of the unphysical spacetime will not be a portion of the Einstein static universe, nor is there any reason to demand that it should be.)

We will accomplish this mapping in three steps.
First, we will show that any FRW universe with non-zero vacuum energy is conformally equivalent to a portion of de Sitter space.
Second, we will show that this portion of de Sitter space can be conformally mapped to a portion of the Einstein static universe whose future boundary is identical to the piece of the future boundary removed from the parent de Sitter diagram.
Third, we show that this portion of the Einstein static universe is conformally equivalent to a new unphysical spacetime which satisfies all of the above properties (and is not a portion of the Einstein static universe).

An open FRW universe, such as the one following the nucleation event, has the metric
\begin{equation} \label{eq-openFRW}
ds_{\rm FRW}^2 = a^2(T) (- d T^2 + d H_3^2)~,
\end{equation}
where $T$ is the conformal time, $a$ is the scale factor, and 
\begin{equation}
d H_3^2 = d \chi^2 +\sinh^2 \chi d \Omega_2^2~.
\end{equation}
All bubble nucleation events are followed by a period of curvature domination in the bubble universe, and hence conformal time is unbounded toward the past.
Furthermore, as long as $\Lambda \neq 0$, conformal time is finite toward the future.
We are thus free to choose the range of $T$ to be $-\infty < T \leq 0$ for every FRW universe under consideration.
The open slicing of de Sitter space (with unit de Sitter length) is one special case of this metric:
\begin{equation}
a_{\rm dS}(T) = \frac{-1}{\sinh T}~.
\end{equation}
Any other FRW universe is conformally equivalent to the de Sitter open slicing, and the conformal factor is given by
\begin{equation}
\Omega_1 = -\frac{1}{a(T) \sinh T}~.
\end{equation}

The second step is to map the open slicing of de Sitter space into a portion of the Einstein static universe whose future boundary is identical to the piece removed from the parent vacuum.
Recall that the Einstein static universe has a metric given by Eq.~\ref{eq-dsmetric1},
\begin{equation}
d\tilde s_0^2 =- d \eta^2 + d\Omega_3^2~,
\end{equation}
and that the Penrose diagram of the parent de Sitter space consists of the region $0>\eta>-\pi$.
We write the $S^3$ metric as
\begin{equation}
d\Omega_3^2 = d\xi^2 + \sin^2\xi d\Omega_2^2~.
\end{equation}
In these coordinates the nucleation event can be taken to be at $\eta =\eta_{\rm nuc}$ and $\xi =0$, so that the portion of future infinity we need to reproduce in the bubble universe is the portion of the round $S^3$ with $0<\xi<-\eta_{\rm nuc}$. \footnote{In this appendix only, we will find it more convenient for $\eta$ to take negative values and increase towards the future boundary.  The $\eta$ defined in the main body of the paper is $\eta_{\rm main}=|\eta_{\rm appendix}|$.}

The open slicing coordinates $(T, \chi)$ are not convenient for this task, so first we will change coordinates to $(\eta_{\rm in}, \xi_{\rm in})$, in terms of which the metric Eq.~\ref{eq-openFRW}, with $a=a_{\rm dS}$, looks like
\begin{equation}\label{eq-inmetric}
ds^2 = \frac{1}{\sin^2 \eta_{\rm in}}\left(-d \eta_{\rm in}^2 + d\Omega_3^2\right)~,
\end{equation}
where we need to ensure that $0<\xi_{\rm in}<-\eta_{\rm nuc}$ when $\eta_{\rm in}=0$.
Then we can simply rescale by $\sin \eta_{\rm in}$ to finish the job.
Now we turn to the task of finding the transformation $(T,\chi)\rightarrow (\eta_{\rm in},\xi_{\rm in})$~.

Let $X_0$, $X_1$, $X_2$, $X_3$, $X_4$ of be the coordinates of 4+1-dimensional Minkowski space, in which de Sitter space is the hyperboloid
\begin{equation}
-X_0^2  + \sum_{i=1}^{3}{X_i^2} + X_4^2 = 1~.
\end{equation}
The relationship between the coordinates $(\eta_{\rm in},\xi_{\rm in})$ and the $X_\mu$ is
\begin{eqnarray} \label{slice_closed}
X_0 &=& \cot \eta_{\rm in} \\
X_i &=& - \frac{1}{\sin \eta_{\rm in}} \sin \xi_{\rm in} \ \hat{n}_i \nonumber \\
X_4 &=& - \frac{1}{\sin \eta_{\rm in}} \cos \xi_{\rm in} \nonumber,
\end{eqnarray}
where $\hat{n}_i$ are unit vectors whose sum equals 1.
We need to specify how the $(\chi,T)$ coordinates of Eq.~\ref{eq-openFRW} relate to the $X_\mu$.
The standard open slicing of de Sitter space is given by
\begin{eqnarray} 
{X_0} &=& - \Ov{ \sinh T} \cosh \chi \\
{X_i} &=& - \Ov{ \sinh T} \sinh \chi \ \hat{n}_i \nonumber \\
{X_4} &=& - \coth T \nonumber~,
\end{eqnarray}
but this is not what we want to do.
We have to remember that the nucleation event is at $\chi=0$, $T=-\infty$, which would be equivalent to $\eta_{\rm in}=-\pi/2$ and $\xi_{\rm in}=0$ if we used this prescription.
To fix the problem, we use the 4+1-dimensional boost symmetry, which is an isometry of the de Sitter space, to move the nucleation event to another position.
Let $\tilde{X}_\mu$ be given by
\begin{eqnarray} \label{boost}
\tilde{X}_0 &=&  \cosh \beta X_0+ \sinh \beta X_4 \\
\tilde{X}_4 &=& \sinh \beta X_0 +  \cosh \beta X_4 \nonumber\\
\tilde{X}_i &=& X_i,~~1\leq i \leq 3~,
\end{eqnarray}
and then define the open coordinates $(T,\chi)$ by
\begin{eqnarray} \label{slice_open}
\tilde{X}_0 &=& - \Ov{ \sinh T} \cosh \chi \\
\tilde{X}_i &=& - \Ov{ \sinh T} \sinh \chi\ \hat{n}_i \nonumber \\
\tilde{X}_4 &=& - \coth T \nonumber~.
\end{eqnarray}
Now the nucleation event is at $\eta_{\rm in}=\eta_{\rm nuc}$ and $\xi_{\rm in}=0$, where $\cot \eta_{\rm nuc} = -\sinh\beta$~.
We can write the relationship between $(T,\chi)$ and $(\eta_{\rm in}, \xi_{\rm in})$ directly in terms of $\eta_{\rm nuc}$ as
\begin{eqnarray}\label{eq-incoords}
\frac{\sinh \chi}{\sinh T} &=& \frac{\sin \xi_{\rm in}}{\sin \eta_{\rm in}} \\
- \cot \eta_{\rm in} &=& \Ov{\sin \etan} \frac{\cosh \chi}{\sinh T} + \cot \etan \coth T~.\nonumber
\end{eqnarray}
Now the metric is in the form Eq.~\ref{eq-inmetric}, and $0<\xi_{\rm in} < -\eta_{\rm nuc}$ when $\eta_{\rm in}=0$.
Hence we can conformally rescale by
\begin{equation}
\Omega_2 = \sin \eta_{\rm in}~,
\end{equation}
to map the bubble universe into a portion of the Einstein static universe with future boundary identical to the piece cut out of the old diagram.

The product $\Omega_1\Omega_2$ succeeds in mapping our original FRW universe to a portion of the Einstein static universe with the correct future boundary.
Now we will act with one final conformal rescaling, $\Omega_3$, which must change the induced unphysical metric along the future light-cone of the nucleation event to match the one in the old de Sitter diagram, thus ensuring that the total conformal transformation is continuous.
We must also demand that $\Omega_3=1$ on the future boundary $\eta_{\rm in}=0$, since we have already fixed that part of the unphysical spacetime.

The induced metric on the future light-cone of the nucleation event has a similar form in both the unphysical bubble coordinates $(\eta_{\rm in}, \xi_{\rm in})$ and the unphysical parent coordinates, which we will now denote as $(\eta_{\rm out}, \xi_{\rm out})$.
In both cases it merely comes from the Einstein static universe metric:
\begin{equation}
ds_{\rm light-cone}^2 = \sin^2\xi_{\rm in/out} \,d\Omega_2^2~,
\end{equation}
where $\xi_{\rm in/out} = \eta_{\rm in/out}- \eta_{\rm nuc}$.
Hence the conformal factor evaluated along the light-cone is
\begin{equation}
\Omega_3 = \frac{\sin \xi_{\rm out}}{\sin\xi_{\rm in}}~.
\end{equation}
We just need to find out how $\xi_{\rm out}$ and $\xi_{\rm in}$ are related.
We can do this by demanding that the induced {\em physical} metrics also be identical on the light-cone.

The induced physical metric on the light-cone from the parent de Sitter space is found by restricting Eq.~\ref{eq-dsmetric} to the relevant surface:
\begin{equation}
ds_0^2 \rightarrow \frac{\sin^2\xi_{\rm out}}{H_0^2 \sin^2 \eta_{\rm out}} d\Omega_2^2~,
\end{equation}
where again $\eta_{\rm out} - \xi_{\rm out} = \eta_{\rm nuc}$.

In terms of the $\chi$ and $T$ coordinates inside the bubble, the relevant light-cone is given by $\chi\rightarrow \infty$, $T\rightarrow -\infty$, $\chi+T = {\rm const}$.
In terms of $\eta_{\rm in}$ and $\xi_{\rm in}$, the light-cone is given by $\eta_{\rm in} - \xi_{\rm in} = \eta_{\rm nuc}$.
One can see that a fixed value of $\chi+T$ is equivalent to a fixed value of $\xi_{\rm in}$.
Since $a(T)\rightarrow Ce^{-T}$ for some constant $C$ as $T\rightarrow -\infty$, we have from Eq.~\ref{eq-openFRW} that
\begin{equation} 
ds_{\rm FRW}^2 \rightarrow C e^{2(\chi+T)} d \Omega_2^2 = C \frac{\sin^2\xi_{\rm in}}{\sin^2\eta_{\rm in}} d \Omega_2^2 ~.
\end{equation}
We must determine the constant $C$, which is related to our convention that $T=0$ on future infinity.
This constant is easily determined by considering the particular light-cone given by $\chi+T=0$.
This is the event horizon of the FRW universe, and its area is given by $A_{\rm EH}$, which depends on the detailed form of $a(T)$.
So $C=A_{\rm EH}/4\pi$.
Finally we have our relationship between the in-coordinates and out-coordinates on the bubble wall hypersurface:
\begin{equation}
\sqrt{\frac{A_{\rm EH}}{4\pi}}\frac{\sin\left(\eta_{\rm in}-\eta_{\rm nuc}\right)}{\sin \eta_{\rm in}} =\frac{1}{H_0} \frac{\sin \left(\eta_{\rm out}-\eta_{\rm nuc}\right)}{\sin \eta_{\rm out}}~.
\end{equation}
Notice that $\eta_{\rm in}$ and $\eta_{\rm out}$ coincide at the nucleation event (both equal to $\eta_{\rm nuc}$) and at future infinity (both equal to zero).

We are free to extend $\Omega_3$ into the rest of the bubble universe in any continuous way we please, so long as it restricts to $1$ on future infinity and $\sin \xi_{\rm out}/\sin\xi_{\rm in}$ on the boundary light-cone.
One might worry about the fact that $\Omega_3$ is multivalued at the point where the bubble wall meets future infinity.
This is not a concern, though, because the only function which need be well defined is the product
\begin{equation}
\Omega = \Omega_1 \Omega_2 \Omega_3 = -\frac{\sin \eta_{\rm in}}{a(T)\sinh T}\Omega_3~.
\end{equation}
In this formula, $\eta_{\rm in} = \eta_{\rm in} (\chi,T)$ as determined by Eqs~\ref{eq-incoords}~.
From here it is easy to see that as we approach the point $\chi=\infty$ along any slice of fixed $T$, including both $T=0$ (future infinity) and $T=-\infty$ (domain wall), we arrive at $\Omega=0$ so long as $\Omega_3$ remains finite. 

Thus we have accomplished our initial task.
The function $\Omega_3$ is very ambiguous, as we have only fixed its behavior at the boundaries of the spacetime.
We can use this additional freedom to make $\Omega$ arbitrarily smooth at the interface between the parent and the bubble universe.
Additionally, one could impose that $\Omega_3$ be identically equal to $1$ not just on the future boundary, but in the entire region $T>T_0$ for some particular $T_0$.
In particular, we can choose $T_0$ to be the time of vacuum domination in the bubble universe.

\bibliographystyle{utcaps}
\bibliography{all}

\providecommand{\href}[2]{#2}\begingroup\raggedright\begin{thebibliography}{10}

\bibitem{BP}
R.~Bousso and J.~Polchinski, ``Quantization of four-form fluxes and dynamical
  neutralization of the cosmological constant,'' {\em JHEP} {\bf 06} (2000)
  006,
\href{http://arxiv.org/abs/hep-th/0004134}{{\tt hep-th/0004134}}.

\bibitem{KKLT}
S.~Kachru, R.~Kallosh, A.~Linde, and S.~P. Trivedi, ``De {S}itter vacua in
  string theory,'' {\em Phys. Rev. D} {\bf 68} (2003)  046005,
\href{http://arxiv.org/abs/hep-th/0301240}{{\tt hep-th/0301240}}.

\bibitem{LinMez93}
A.~Linde and A.~Mezhlumian, ``Stationary universe,'' {\em Phys. Lett.} {\bf
  B307} (1993)  25--33, \href{http://arxiv.org/abs/gr-qc/9304015}{{\tt
  gr-qc/9304015}}.

\bibitem{LinLin94}
A.~Linde, D.~Linde, and A.~Mezhlumian, ``From the Big Bang theory to the theory
  of a stationary universe,'' {\em Phys. Rev. D} {\bf 49} (1994)  1783--1826,
  \href{http://arxiv.org/abs/gr-qc/9306035}{{\tt gr-qc/9306035}}.

\bibitem{GarLin94}
J.~Garc{\'{\i}}a-Bellido, A.~Linde, and D.~Linde, ``Fluctuations of the
  gravitational constant in the inflationary {B}rans-{D}icke cosmology,'' {\em
  Phys. Rev. D} {\bf 50} (1994)  730--750,
  \href{http://arxiv.org/abs/astro-ph/9312039}{{\tt astro-ph/9312039}}.

\bibitem{GarLin94a}
J.~Garc{\'{\i}}a-Bellido and A.~D. Linde, ``Stationarity of inflation and
  predictions of quantum cosmology,'' {\em Phys. Rev.} {\bf D51} (1995)
  429--443, \href{http://arxiv.org/abs/hep-th/9408023}{{\tt hep-th/9408023}}.

\bibitem{GarLin95}
J.~Garc{\'{\i}}a-Bellido and A.~Linde, ``Stationary solutions in
  {B}rans-{D}icke stochastic inflationary cosmology,'' {\em Phys. Rev. D} {\bf
  52} (1995)  6730--6738, \href{http://arxiv.org/abs/gr-qc/9504022}{{\tt
  gr-qc/9504022}}.

\bibitem{GarSch05}
J.~Garriga, D.~Schwartz-Perlov, A.~Vilenkin, and S.~Winitzki, ``Probabilities
  in the inflationary multiverse,'' {\em JCAP} {\bf 0601} (2006)  017,
\href{http://arxiv.org/abs/hep-th/0509184}{{\tt hep-th/0509184}}.

\bibitem{VanVil06}
V.~Vanchurin and A.~Vilenkin, ``Eternal observers and bubble abundances in the
  landscape,''
\href{http://arxiv.org/abs/hep-th/0605015}{{\tt hep-th/0605015}}.

\bibitem{Van06}
V.~Vanchurin, ``Geodesic measures of the landscape,'' {\em Phys. Rev. D} {\bf
  75} (2007)  023524,
\href{http://arxiv.org/abs/hep-th/0612215}{{\tt hep-th/0612215}}.

\bibitem{Bou06}
R.~Bousso, ``Holographic probabilities in eternal inflation,'' {\em Phys. Rev.
  Lett.} {\bf 97} (2006)  191302,
\href{http://arxiv.org/abs/hep-th/0605263}{{\tt hep-th/0605263}}.

\bibitem{Lin06}
A.~Linde, ``Sinks in the Landscape, {B}oltzmann {B}rains, and the Cosmological
  Constant Problem,'' {\em JCAP} {\bf 0701} (2007)  022,
\href{http://arxiv.org/abs/hep-th/0611043}{{\tt hep-th/0611043}}.

\bibitem{Lin07}
A.~Linde, ``Towards a gauge invariant volume-weighted probability measure for
  eternal inflation,'' {\em JCAP} {\bf 0706} (2007)  017,
\href{http://arxiv.org/abs/arXiv:0705.1160 [hep-th]}{{\tt arXiv:0705.1160
  [hep-th]}}.

\bibitem{Pag08}
D.~N. Page, ``{Cosmological Measures without Volume Weighting},''
  \href{http://dx.doi.org/10.1088/1475-7516/2008/10/025}{{\em JCAP} {\bf 0810}
  (2008)  025},
\href{http://arxiv.org/abs/0808.0351}{{\tt arXiv:0808.0351 [hep-th]}}.

\bibitem{GarVil08}
J.~Garriga and A.~Vilenkin, ``{Holographic Multiverse},''
  \href{http://dx.doi.org/10.1088/1475-7516/2009/01/021}{{\em JCAP} {\bf 0901}
  (2009)  021},
\href{http://arxiv.org/abs/0809.4257}{{\tt arXiv:0809.4257 [hep-th]}}.

\bibitem{GarVil09}
  J.~Garriga and A.~Vilenkin,
  ``Holographic multiverse and conformal invariance,''
  JCAP {\bf 0911}, 020 (2009)
  [arXiv:0905.1509 [hep-th]].


\bibitem{Win08a}
S.~Winitzki, ``{A volume-weighted measure for eternal inflation},''
  \href{http://dx.doi.org/10.1103/PhysRevD.78.043501}{{\em Phys. Rev.} {\bf
  D78} (2008)  043501}, \href{http://arxiv.org/abs/0803.1300}{{\tt
  arXiv:0803.1300 [gr-qc]}}.

\bibitem{Win08b}
S.~Winitzki, ``{Reheating-volume measure for random-walk inflation},''
  \href{http://dx.doi.org/10.1103/PhysRevD.78.063517}{{\em Phys. Rev.} {\bf
  D78} (2008)  063517}, \href{http://arxiv.org/abs/0805.3940}{{\tt
  arXiv:0805.3940 [gr-qc]}}.

\bibitem{Win08c}
S.~Winitzki, ``{Reheating-volume measure in the landscape},''
  \href{http://dx.doi.org/10.1103/PhysRevD.78.123518}{{\em Phys. Rev.} {\bf
  D78} (2008)  123518}, \href{http://arxiv.org/abs/0810.1517}{{\tt
  arXiv:0810.1517 [gr-qc]}}.

\bibitem{LinVan08}
A.~Linde, V.~Vanchurin, and S.~Winitzki, ``{Stationary Measure in the
  Multiverse},'' \href{http://dx.doi.org/10.1088/1475-7516/2009/01/031}{{\em
  JCAP} {\bf 0901} (2009)  031}, \href{http://arxiv.org/abs/0812.0005}{{\tt
  arXiv:0812.0005 [hep-th]}}.

\bibitem{Bou09}
R.~Bousso, ``{Complementarity in the Multiverse},''
  \href{http://dx.doi.org/10.1103/PhysRevD.79.123524}{{\em Phys. Rev.} {\bf
  D79} (2009)  123524},
\href{http://arxiv.org/abs/0901.4806}{{\tt arXiv:0901.4806 [hep-th]}}.

\bibitem{GarVil05}
J.~Garriga and A.~Vilenkin, ``Anthropic prediction for {L}ambda and the {Q}
  catastrophe,'' {\em Prog. Theor. Phys. Suppl.} {\bf 163} (2006)  245--257,
\href{http://arxiv.org/abs/hep-th/0508005}{{\tt hep-th/0508005}}.

\bibitem{FelHal05}
B.~Feldstein, L.~J. Hall, and T.~Watari, ``Density perturbations and the
  cosmological constant from inflationary landscapes,'' {\em Phys. Rev. D} {\bf
  72} (2005)  123506,
\href{http://arxiv.org/abs/hep-th/0506235}{{\tt hep-th/0506235}}.

\bibitem{PogVil06}
L.~Pogosian and A.~Vilenkin, ``{Anthropic predictions for vacuum energy and
  neutrino masses in the light of WMAP-3},'' {\em JCAP} {\bf 0701} (2007)  025,
\href{http://arxiv.org/abs/astro-ph/0611573}{{\tt arXiv:astro-ph/0611573}}.

\bibitem{FelHal06}
B.~Feldstein, L.~J. Hall, and T.~Watari, ``{Landscape Predictions for the Higgs
  Boson and Top Quark Masses},''
  \href{http://dx.doi.org/10.1103/PhysRevD.74.095011}{{\em Phys. Rev.} {\bf
  D74} (2006)  095011},
\href{http://arxiv.org/abs/hep-ph/0608121}{{\tt arXiv:hep-ph/0608121}}.

\bibitem{SchVil06}
D.~Schwartz-Perlov and A.~Vilenkin, ``Probabilities in the
  {B}ousso-{P}olchinski multiverse,'' {\em JCAP} {\bf 0606} (2006)  010,
\href{http://arxiv.org/abs/hep-th/0601162}{{\tt hep-th/0601162}}.

\bibitem{Sch06}
D.~Schwartz-Perlov, ``Probabilities in the
  {A}rkani-{H}amed-{D}imopolous-{K}achru landscape,''
\href{http://arxiv.org/abs/hep-th/0611237}{{\tt hep-th/0611237}}.

\bibitem{BouHar07}
R.~Bousso, R.~Harnik, G.~D. Kribs, and G.~Perez, ``Predicting the cosmological
  constant from the causal entropic principle,'' {\em Phys. Rev. D} {\bf 76}
  (2007)  043513,
\href{http://arxiv.org/abs/hep-th/0702115}{{\tt hep-th/0702115}}.

\bibitem{BouYan07}
R.~Bousso and I.-S. Yang, ``Landscape Predictions from Cosmological Vacuum
  Selection,'' {\em Phys. Rev. D} {\bf 75} (2007)  123520,
\href{http://arxiv.org/abs/hep-th/0703206}{{\tt hep-th/0703206}}.

\bibitem{OluSch07}
K.~D. Olum and D.~Schwartz-Perlov, ``Anthropic prediction in a large toy
  landscape,''
\href{http://arxiv.org/abs/arXiv:0705.2562 [hep-th]}{{\tt arXiv:0705.2562
  [hep-th]}}.

\bibitem{BouFre07}
R.~Bousso, B.~Freivogel, and I.-S. Yang, ``{Boltzmann babies in the proper time
  measure},'' \href{http://dx.doi.org/10.1103/PhysRevD.77.103514}{{\em Phys.
  Rev.} {\bf D77} (2008)  103514},
\href{http://arxiv.org/abs/0712.3324}{{\tt arXiv:0712.3324 [hep-th]}}.

\bibitem{Sch08}
D.~Schwartz-Perlov, ``{Anthropic prediction for a large multi-jump
  landscape},'' \href{http://dx.doi.org/10.1088/1475-7516/2008/10/009}{{\em
  JCAP} {\bf 0810} (2008)  009},
\href{http://arxiv.org/abs/0805.3549}{{\tt arXiv:0805.3549 [hep-th]}}.

\bibitem{BouLei08}
R.~Bousso and S.~Leichenauer, ``{Star Formation in the Multiverse},''
\href{http://arxiv.org/abs/0810.3044}{{\tt arXiv:0810.3044 [astro-ph]}}.

\bibitem{Pag06b}
D.~N. Page, ``Return of the {B}oltzmann brains,''
\href{http://arxiv.org/abs/hep-th/0611158}{{\tt hep-th/0611158}}.

\bibitem{HalNom07}
L.~J. Hall and Y.~Nomura, ``{Evidence for the Multiverse in the Standard Model
  and Beyond},'' \href{http://dx.doi.org/10.1103/PhysRevD.78.035001}{{\em Phys.
  Rev.} {\bf D78} (2008)  035001},
\href{http://arxiv.org/abs/0712.2454}{{\tt arXiv:0712.2454 [hep-ph]}}.

\bibitem{HalSal07a}
L.~J. Hall, M.~P. Salem, and T.~Watari, ``{Statistical Understanding of Quark
  and Lepton Masses in Gaussian Landscapes},''
  \href{http://dx.doi.org/10.1103/PhysRevD.76.093001}{{\em Phys. Rev.} {\bf
  D76} (2007)  093001},
\href{http://arxiv.org/abs/0707.3446}{{\tt arXiv:0707.3446 [hep-ph]}}.

\bibitem{HalSal07b}
L.~J. Hall, M.~P. Salem, and T.~Watari, ``{Quark and Lepton Masses from
  Gaussian Landscapes},''
  \href{http://dx.doi.org/10.1103/PhysRevLett.100.141801}{{\em Phys. Rev.
  Lett.} {\bf 100} (2008)  141801},
\href{http://arxiv.org/abs/0707.3444}{{\tt arXiv:0707.3444 [hep-ph]}}.

\bibitem{DGSV08}
A.~De~Simone, A.~H. Guth, M.~P. Salem, and A.~Vilenkin, ``{Predicting the
  cosmological constant with the scale-factor cutoff measure},''
\href{http://arxiv.org/abs/0805.2173}{{\tt arXiv:0805.2173 [hep-th]}}.

\bibitem{BouFre08b}
R.~Bousso, B.~Freivogel, and I.-S. Yang, ``{Properties of the scale factor
  measure},''
\href{http://arxiv.org/abs/0808.3770}{{\tt arXiv:0808.3770 [hep-th]}}.

\bibitem{DGLNSV08}
A.~De~Simone {\em et al.}, ``{Boltzmann brains and the scale-factor cutoff
  measure of the multiverse},''
\href{http://arxiv.org/abs/0808.3778}{{\tt arXiv:0808.3778 [hep-th]}}.

\bibitem{HalSal08}
L.~J. Hall, M.~P. Salem, and T.~Watari, ``{Neutrino mixing and mass hierarchy
  in Gaussian landscapes},''
  \href{http://dx.doi.org/10.1103/PhysRevD.79.025010}{{\em Phys. Rev.} {\bf
  D79} (2009)  025010},
\href{http://arxiv.org/abs/0810.2561}{{\tt arXiv:0810.2561 [hep-th]}}.

\bibitem{Sal09}
M.~P. Salem, ``{Negative vacuum energy densities and the causal diamond
  measure},''
\href{http://arxiv.org/abs/0902.4485}{{\tt arXiv:0902.4485 [hep-th]}}.

\bibitem{BouHal09}
R.~Bousso, L.~J. Hall, and Y.~Nomura, ``{Multiverse Understanding of
  Cosmological Coincidences},''
  \href{http://dx.doi.org/10.1103/PhysRevD.80.063510}{{\em Phys. Rev.} {\bf
  D80} (2009)  063510},
\href{http://arxiv.org/abs/0902.2263}{{\tt arXiv:0902.2263 [hep-th]}}.

\bibitem{PhiAlb09}
D.~Phillips and A.~Albrecht, ``{Effects of Inhomogeneity on the Causal Entropic
  prediction of Lambda},''
\href{http://arxiv.org/abs/0903.1622}{{\tt arXiv:0903.1622 [gr-qc]}}.

\bibitem{BouLei09}
R.~Bousso and S.~Leichenauer, ``{Predictions from Star Formation in the
  Multiverse},''
\href{http://arxiv.org/abs/0907.4917}{{\tt arXiv:0907.4917 [hep-th]}}.

\bibitem{HalNom09}
L.~J. Hall and Y.~Nomura, ``{A Finely-Predicted Higgs Boson Mass from A
  Finely-Tuned Weak Scale},''
  \href{http://dx.doi.org/10.1007/JHEP03(2010)076}{{\em JHEP} {\bf 03} (2010)
  076},
\href{http://arxiv.org/abs/0910.2235}{{\tt arXiv:0910.2235 [hep-ph]}}.

\bibitem{EloGoh09}
G.~Elor, H.-S. Goh, L.~J. Hall, P.~Kumar, and Y.~Nomura, ``{Environmentally
  Selected WIMP Dark Matter with High-Scale Supersymmetry Breaking},''
\href{http://arxiv.org/abs/0912.3942}{{\tt arXiv:0912.3942 [hep-ph]}}.

\bibitem{SusWit98}
L.~Susskind and E.~Witten, ``The holographic bound in {A}nti-de~{S}itter
  space,'' \href{http://arxiv.org/abs/{h}ep-th/9805114}{{\tt
  {h}ep-th/9805114}}.

\bibitem{BouRan01}
R.~Bousso and L.~Randall, ``Holographic domains of {A}nti-de {S}itter space,''
  {\em JHEP} {\bf 04} (2002)  057,
\href{http://arXiv.org/abs/hep-th/0112080}{{\tt hep-th/0112080}}.

\bibitem{HawEll}
S.~W. Hawking and G.~F.~R. Ellis, {\em The large scale stucture of space-time}.
\newblock Cambridge University Press, Cambridge, England, 1973.

\bibitem{Scho84}
R.~Schoen, ``Conformal deformation of a Riemannian metric to constant scalar
  curvature,'' {\em J. Diff. Geom.} {\bf 20} (1984)  479--495.

\bibitem{Lee87}
J.~M. Lee and T.~H. Parker, ``The Yamabe Problem,'' {\em Bull Amer. Math. Soc.}
  {\bf 17} (1987)  37--81.

\bibitem{Yam60}
H.~Yamabe, ``On a deformation of Riemannian structures on compact manifolds,''
  {\em Osaka Math. J.} {\bf 12} (1960)  21--37.

\bibitem{Tru68}
N.~Trudinger, ``Remarks concerning the conformal deformation of Riemannian
  structures on compact manifolds,'' {\em Ann. Scuola Norm. Sup. Pisa} {\bf 22}
  (1968)  265--274.

\bibitem{Aub76}
T.~Aubin, ``Equations diff\'erentielles non lin\'eaires et probl\`eme de Yamabe
  concernant la courbure scalaire,'' {\em J. Math. Pures Appl.} {\bf 55} (1976)
   269--296.

\bibitem{And05}
M.~Anderson, ``On uniqueness and differentiability in the space of Yamabe
  metrics,'' {\em Comm. Contemp. Math.} {\bf 7} (2005)  299--310,
  \href{http://arxiv.org/abs/math/0210446v3 [math.DG]}{{\tt math/0210446v3
  [math.DG]}}.

\bibitem{BouYan09}
R.~Bousso and I.-S. Yang, ``{Global-Local Duality in Eternal Inflation},''
  \href{http://dx.doi.org/10.1103/PhysRevD.80.124024}{{\em Phys. Rev.} {\bf
  D80} (2009)  124024},
\href{http://arxiv.org/abs/0904.2386}{{\tt arXiv:0904.2386 [hep-th]}}.

\bibitem{GerPen72}
R.~P. Geroch, E.~H. Kronheimer, and R.~Penrose, ``{Ideal points in
  space-time},''
{\em Proc. Roy. Soc. Lond.} {\bf A327} (1972)  545--567.

\bibitem{BouGib82}
W.~Boucher and G.~W. Gibbons, ``{Cosmic Baldness},''. Presented at 1982
  Nuffield Workshop on the Very Early Universe, Cambridge, England, Jun 21 -
  Jul 9, 1982.

\bibitem{BouHar10}
R.~Bousso and R.~Harnik, ``{The Entropic Landscape},''
\href{http://arxiv.org/abs/1001.1155}{{\tt arXiv:1001.1155 [hep-th]}}.

\bibitem{Wei87}
S.~Weinberg, ``ANTHROPIC BOUND ON THE COSMOLOGICAL CONSTANT,''
{\em Phys. Rev. Lett.} {\bf 59} (1987)  2607.

\bibitem{Dys79}
F.~Dyson, ``TIME WITHOUT END: {P}HYSICS AND BIOLOGY IN AN OPEN UNIVERSE,''
{\em Rev. Mod. Phys.} {\bf 51} (1979)  447--460.

\end{thebibliography}\endgroup
\end{document}